\documentclass[prd,nofootinbib]{revtex4}

\usepackage{amsmath,amssymb}
\usepackage{mathrsfs}
\usepackage{mathtools}
\usepackage{rotating}
\usepackage{mathrsfs}
\usepackage{pstricks}
\usepackage{pst-func}
\usepackage{graphicx}

\def\dbl{\hbox{${1\hskip -2.2pt{\rm l}}$}}

\begin{document}

\title{Oversights in the Respective Theorems of von Neumann and Bell are Homologous}

\author{Joy Christian}

\email{jjc@bu.edu}

\affiliation{Einstein Centre for Local-Realistic Physics, 15 Thackley End, Oxford OX2 6LB, United Kingdom}

\begin{abstract}
We show that the respective oversights in the von Neumann's general theorem against all hidden variable theories and Bell's theorem against their local-realistic counterparts are homologous. Both theorems unjustifiably assume the additivity of expectation values within hidden variable theories to derive their respective conclusions. However, for non-commuting observables, the equivalence of a sum of expectation values and the expectation value of the sum of measurement results, although respected within quantum mechanics, need not hold for hidden variable theories, regardless of specific characteristics such as local realism they may respect. Once this oversight is ameliorated from Bell's argument and local realism is implemented correctly, the bounds on the CHSH correlator work out to be ${\pm2\sqrt{2}}$ instead of ${\pm2}$, thereby mitigating the conclusion of Bell's theorem. Consequently, what is ruled out by the Bell-test experiments is not local realism but the additivity of expectation values.
\end{abstract}

\maketitle

\parskip 4.7pt

\baselineskip 10.3pt

\section{Introduction}

One of the characteristic features of quantum theory is the fact that the result of any individual measurement event is not determined by the theory. If the quantum state of a physical system is not represented by an eigenvector of the observable being measured, then the theory predicts only a statistical distribution of possible results of measurements. This apparent deficiency --- and it can be viewed as deficiency --- has long inspired investigations into the possibility of a deeper theory underlying quantum theory, built on the so-called ``dispersion-free" states \cite{vonNeumann}. Such states are supposed to be specified by ``hidden variables", in addition to the usual quantum mechanical state vector, and determine the outcomes of individual measurement events. That, in turn, implies simultaneous assignment of definite values to all of the observables of a quantum mechanical system, whether or not they are actually observed, making the resulting theory compatible with Einstein's conception of realism \cite{EPR,Howard}. The hypothetical variables accomplishing this feat are called ``hidden" because if states with prescribed values of such variables can be actually prepared, then quantum theory would be rendered observationally inadequate. 

Against this background, von Neumann set out to prove in his theorem \cite{vonNeumann}, based on a set of four assumptions \cite{Albertson,Ballentine},\break that no hidden variable theories consisting dispersion-free states are possible that could simultaneously assign definite values to all the observables of a quantum system consistently \cite{Shimony}. In other words, he set out to prove that any theory supplementing quantum theory with hidden variables to produce dispersion-free states would be logically inconsistent.

In recent decades Bell's theorem \cite{Bell-1964} has become much more familiar compared to von Neumann's theorem because of its significance for quantum information theory, quantum computing, quantum cryptography, and other quantum technologies. But unlike von Neumann's theorem, it does not purport to rule out all hidden variable theories. In fact, Bell was a supporter of Bohm's non-local hidden variable theory \cite{Bohm}. Indeed, the scope of Bell's theorem is limited. It claims that no locally causal and realistic theory of the kind envisaged by Einstein and formed the conceptual basis for the argument of Einstein, Podolsky, and Rosen \cite{EPR} can reproduce all of the statistical predictions of quantum theory. 

By now it is well known that von Neumann's theorem against hidden variable theories reproducing the predictions of quantum theory contains a rather serious oversight \cite{Hermann,Jauch,Bell-1966,Siegel,Kochen,Jammer,Guido}. What is less well known and less appreciated in some quarters is that Bell's theorem against {\it local} hidden variable theories also contains a similar oversight, in addition to other questionable assumptions \cite{Disproof,IJTP,Bell-oversight,RSOS,IEEE-Access,Socks}. In what follows, we compare the respective oversights in the theorems of von Neumann and Bell and show that they are homologous. It is important to bear in mind, however, that while Bell's theorem is not a theorem within quantum theory itself, von Neumann's theorem is a mathematically correct theorem providing significant other results for the logical foundations of quantum theory. But both theorems assume the additivity of expectation values to derive their respective conclusions. However, for non-commuting observables, the equivalence of any sum of expectation values and the expectation value of the sum, although respected within quantum mechanics, need not hold within hidden variable theories, regardless of specific characteristics such as locality or realism they may be required to respect. Once this oversight is ameliorated from Bell's argument, the bounds on the CHSH correlator work out to be ${\pm2\sqrt{2}}$ instead of ${\pm2}$, thereby mitigating the conclusion of Bell's theorem. This, in turn, implies that what is ruled out by the Bell-test experiments is not local realism as widely believed, but the additivity of expectation values, which does not hold in general for any hidden variable theories to begin with. 

\section{The Oversight in von Neumann's Theorem }\label{II}

It appears that Einstein was aware of the oversight in von Neumann's theorem before the publication of his argument against the completeness of quantum mechanics in 1935, co-authored with Podolsky and Rosen \cite{EPR}. He, however, does not seem to have published his criticism anywhere. We learn about it from what has been reported by Shimony some six\break decades later in the collection of his own papers \cite{Shimony}. In the third paragraph of Section 1 of Chapter 7, Shimony writes:

\begin{quote}
The great paper of Einstein, Podolsky and Rosen \cite{EPR} of 1935 concludes that quantum mechanics is an incomplete theory, without suggesting that changes of the theory are prerequisites to the job of completion. They make no reference to von Neumann's argument, which had been published three years earlier, although they could hardly have been unaware of it. ... It is conceivable that Einstein was critical of one or more of von Neumann's premises, as were Bell \cite{Bell-1966}, Siegel \cite{Siegel}, and Kochen and Specker \cite{Kochen} thirty years later. However, in the absence of specific evidence that this is so, I doubt it.    
\end{quote}
As noted, this quote is from the Chapter 7 in the collection of Shimony's papers, which was originally published in 1971 \cite{Shimony}. But by 1993 Shimony's doubt about Einstein's awareness and criticism of von Neumann's oversight were put to rest\break by Peter G. Bergmann, who was Einstein's collaborator. In the comment added to his original paper, Shimony writes: 
\begin{quote}
The question raised in the third paragraph of Section 1 was cleared up by a conversation with Prof. Peter G. Bergmann around 1980. He recalled a discussion with Einstein and Valentine Bargmann around 1938 at the Institute for Advanced Study, during which Einstein took von Neumann's book from the shelf and pointed to premise B$'$ of von Neumann's theorem (in Section 1 of Chapter IV): ``If $\,\mathcal R$, $\mathcal S$, $\dots$ are arbitrary quantities and $a$, $b$, $\dots$ real numbers, then Exp$(a\,{\mathcal R} + b\,{\mathcal S} + \dots) =$ $a$\,Exp$({\mathcal R}) + b\,$Exp$({\mathcal S}) + \dots\,$." Einstein then said that there is no reason why this premise should hold in a state not acknowledged by quantum mechanics if $\mathcal R$, $\mathcal S$, etc. are not simultaneously measurable. Einstein's criticism is essentially the same as those of Siegel \cite{Siegel}, Jauch and Piron \cite{Jauch}, Bell \cite{Bell-1966}, and Kochen and Specker \cite{Kochen} nearly thirty years later.  
\end{quote}
Although Shimony does not mention Grete Hermann's contribution to the debate \cite{Hermann}, by now it is well known \cite{Jammer,Mermin1993,Guido} that she was the first to publish a criticism of von Neumann's argument in her 1935 essay on the philosophy of quantum mechanics, essentially along the line of the criticism by Einstein, Bell \cite{Bell-1966}, Siegel \cite{Siegel}, and Kochen and Specker \cite{Kochen}. As we shall see in more detail below, the essence of Einstein's argument is that, for non-commuting observables, the assumed equivalence of a sum of expectation values and the expectation value of the sum of corresponding measurement results, although respected by quantum mechanical expectation values, need not hold within hidden variable theories, regardless of whether those theories are local or nonlocal.

Among the criticisms of von Neumann's assumption, the one that appears to have struck the chord is that by Bell \cite{Bell-1966}. In anticipation of what we wish to demonstrate below, let us restrict his assumption to four arbitrary observables, ${\mathcal R}$, ${\mathcal S}$, ${\mathcal T}$, and ${\mathcal U}$, and four real numbers, $a$, $b$, $c$, and $d$. Then, within quantum theory, using normalized state vectors ${|\,\psi\rangle}$,\break his assumption is: Given the observables ${\mathcal R}$, ${\mathcal S}$, ${\mathcal T}$, and ${\mathcal U}$, there exists an observable $a\,{\mathcal R} + b\,{\mathcal S} + c\,{\mathcal T} + d\;{\mathcal U}$ such that
\begin{equation}
a \left\langle\psi\left|\,{\mathcal R}\,\right|\psi\right\rangle + b\left\langle\psi\left|\,{\mathcal S}\,\right|\psi\right\rangle + c\left\langle\psi\left|\,{\mathcal T}\,\right|\psi\right\rangle + d\left\langle\psi\left|\,{\mathcal U}\,\right|\psi\right\rangle\,=\,\left\langle\psi\left|\left\{a\,{\mathcal R} + b\,{\mathcal S} + c\,{\mathcal T} + d\;{\mathcal U\,}\right\}\right|\psi\right\rangle,\label{QM}
\end{equation}
where ${\left\langle\psi\left|\,{\mathcal R}\,\right|\psi\right\rangle}$ represents the quantum mechanical expectation value of the self-adjoint operator ${\mathcal R}$ in the state $|\psi\rangle$. His theorem then asserts that there exists a self-adjoint operator ${\mathcal R}$ on the corresponding Hilbert space ${\mathscr H}$ such that
\begin{equation}
{\left\langle\psi\left|\,{\mathcal R}^2\,\right|\psi\right\rangle}\not={\left\langle\psi\left|\,{\mathcal R}\,\right|\psi\right\rangle}^2.
\end{equation}
That is to say, the state defined by ${\left\langle\psi\left|\,{\mathcal R}\,\right|\psi\right\rangle}$ is not dispersion-free over the quantum mechanical observables ${\mathcal R}$. But as Bell points out in his exposition of von Neumann's oversight \cite{Bell-1966}, the relation (\ref{QM}) --- {\it i.e.}, the additivity of expectation values --- is one of the peculiar properties of quantum mechanical states. It need not hold in a hidden variable theory.

It is not difficult to understand why von Neumann was led to assume otherwise. With a slight abuse of notation, we may represent a hypothetical dispersion-free state by a unit vector $\left|\,\psi,\,\lambda\,\right\rangle$ characterizing a hidden variable theory that\break has no statistical character by definition, in contrast to a quantum theory with a genuine unit vector $\left|\psi\right\rangle$ in a Hilbert space. Then, the expectation value of any observable $\mathcal{R}$ would be necessarily equal to one of its eigenvalues ${\mathscr R}({\lambda_k})$:
\begin{equation}
\left\langle\,\psi,\,\lambda_k\,|\,\mathcal{R}\,|\,\psi,\,\lambda_k\,\right\rangle = {\mathscr R}({\lambda_k}),
\end{equation}
where ${\mathscr R}({\lambda_k})$ is a unique scalar determined by a specific hidden variable ${\lambda_k\in\{\lambda_1, \dots,\lambda_n\}}$ and the eigenvalue equation
\begin{equation}
\mathcal{R}\left|\,\psi\,\right\rangle = {\mathscr R} \left|\,\psi\,\right\rangle\!,
\end{equation}
so that the measurement of $\mathcal{R}$ in the state $\left|\,\psi,\,\lambda\,\right\rangle$ specified by $\psi$ and $\lambda$ results with certainty in that eigenvalue ${\mathscr R}({\lambda_k})$. Consequently, such a dispersion-free state would satisfy
\begin{equation}
\left\langle\,\psi,\,\lambda\,|(\,\mathcal{R}-\left\langle\,\psi,\,\lambda\,|\,\mathcal{R}\,|\,\psi,\,\lambda\,\right\rangle)^2|\,\psi,\,\lambda\,\right\rangle=0,
\end{equation}
or, equivalently,
\begin{equation}
\left\langle\,\psi,\,\lambda\,|\,\mathcal{R}^2\,|\,\psi,\,\lambda\,\right\rangle = \left\langle\,\psi,\,\lambda\,|\,\mathcal{R}\,|\,\psi,\,\lambda\,\right\rangle^2\!,
\end{equation}
for all observables $\mathcal{R}$, since --- by hypothesis --- every physical quantity would have a unique value ${\mathscr R}({\lambda_k})$. The quantum mechanical state specified by $|\psi\rangle$ is then recovered by uniform averaging over $\lambda$, which would give the expectation value
\begin{equation}
\left\langle\,\psi\,|\,\mathcal{R}\,|\,\psi\,\right\rangle =\lim_{\,n\,\gg\,1} \frac{1}{n}\sum_{k\,=\,1}^{n}
\left\langle\,\psi,\,\lambda_k\,|\,\mathcal{R}\,|\,\psi,\,\lambda_k\,\right\rangle = \lim_{\,n\,\gg\,1}\frac{1}{n}\sum_{k\,=\,1}^{n}
{\mathscr R}({\lambda}_k). \label{7}
\end{equation}
In what follows we will omit the limits on summations for convenience, with the large $n$ limits understood implicitly.

It is important to note that Eq.~(\ref{7}) exhibits an inter-theoretical relation between quantities pertaining to two entirely different theories. On its LHS is the standard quantum mechanical expectation value of the operator ${\mathcal R}$ in the state $|\,\psi\,\rangle$, and on its RHS is a classical average of ${\mathscr R}(\lambda_k)$ over the hidden variables ${\lambda}$, with no physical significance attributed to $\lambda$. To be sure, a fully developed hidden variable theory would have to include a detailed theory of measurement accounting for the behaviour of $\lambda$ during the measurement process itself, but such a detailed theory is not necessary for our analysis. On the other hand, as noted in the Introduction, if physical states supplemented with such hidden variables can be actually prepared, then quantum theory would be rendered observationally inadequate. Therefore, $\lambda$ must be traced out by uniform averaging, as done in Eq.~(\ref{7}). An explicit example of how that works can be found in Section 2 of Bell's paper \cite{Bell-1966}.

Since by hypothesis the inter-theoretical relation (\ref{7}) holds for all observables, we can express the LHS of Eq.~(\ref{QM}) as
\begin{equation}
a \left\langle\psi\left|\,{\mathcal R}\,\right|\psi\right\rangle + b\left\langle\psi\left|\,{\mathcal S}\,\right|\psi\right\rangle + c\left\langle\psi\left|\,{\mathcal T}\,\right|\psi\right\rangle + d\left\langle\psi\left|\,{\mathcal U}\,\right|\psi\right\rangle=
\frac{a}{n}\!\sum_{k\,=\,1}^{n}
{\mathscr R}({\lambda}_k)+
\frac{b}{n}\!\sum_{k\,=\,1}^{n}
{\mathscr S}({\lambda}_k)+
\frac{c}{n}\!\sum_{k\,=\,1}^{n}
{\mathscr T}({\lambda}_k)+
\frac{d}{n}\!\sum_{k\,=\,1}^{n}
{\mathscr U}({\lambda}_k)\,,
\end{equation}
where each of the eigenvalues ${\mathscr R}({\lambda}_k)$, ${\mathscr S}({\lambda}_k)$, ${\mathscr T}({\lambda}_k)$, and ${\mathscr U}({\lambda}_k)$ is a unique scalar number. But that makes the last equality a slippery slope. Because every schoolchild knows that a sum of averages is equal to the average of the sum:
\begin{equation}
\frac{a}{n}\sum_{k\,=\,1}^{n}
{\mathscr R}({\lambda}_k)+
\frac{b}{n}\sum_{k\,=\,1}^{n}
{\mathscr S}({\lambda}_k)+
\frac{c}{n}\sum_{k\,=\,1}^{n}
{\mathscr T}({\lambda}_k)+
\frac{d}{n}\sum_{k\,=\,1}^{n}
{\mathscr U}({\lambda}_k)=
\frac{1}{n}\sum_{k\,=\,1}^{n}
\big\{a\,{\mathscr R}({\lambda}_k) + b\,{\mathscr S}({\lambda}_k) + c\,{\mathscr T}({\lambda}_k) + d\,{\mathscr U}({\lambda}_k)\big\}. \label{CM}
\end{equation}

Comparing the equalities (\ref{QM}) and (\ref{CM}) it is indeed tempting to conclude that (\ref{CM}) is a natural dispersion-free counterpart of the quantum mechanical property (\ref{QM}), and therefore additivity of expectation values must hold also in any hidden variable theory, as von Neumann had assumed. But the obvious disanalogy between (\ref{QM}) and (\ref{CM}) is that, while ${\mathcal R}$, ${\mathcal S}$, ${\mathcal T}$, and ${\mathcal U}$ are Hermitian operators in a complex Hilbert space and therefore may not commute with each other in general, ${\mathscr R}({\lambda}_k)$, ${\mathscr S}({\lambda}_k)$, ${\mathscr T}({\lambda}_k)$, and ${\mathscr U}({\lambda}_k)$ --- being the specific results of the measurements of the observables ${\mathcal R}$, ${\mathcal S}$, ${\mathcal T}$, and ${\mathcal U}$ --- are scalar numbers and therefore necessarily commute with each other. More importantly, as explained by Bell \cite{Bell-1966}, eigenvalues of non-commuting observables are not additive \cite{Hermann}. The result of the measurement of the sum $a\,{\mathcal R} + b\,{\mathcal S} + c\,{\mathcal T} + d\;{\mathcal U}$ of non-commuting operators will not be the sum $a\,{\mathscr R}({\lambda}_k) + b\,{\mathscr S}({\lambda}_k) + c\,{\mathscr T}({\lambda}_k) + d\,{\mathscr U}({\lambda}_k)$ of the\break eigenvalues of those operators in a dispersion-free state. Consequently, for non-commuting operators ${\mathcal R}$, ${\mathcal S}$, ${\mathcal T}$, and ${\mathcal U}$,
\begin{equation}
\left\langle\,\psi,\,\lambda_k\left|\left\{a\,{\mathcal R} + b\,{\mathcal S} + c\,{\mathcal T} + d\;{\mathcal U\,}\right\}\right|\psi,\,\lambda_k\,\right\rangle \not=
a\,{\mathscr R}({\lambda}_k) + b\,{\mathscr S}({\lambda}_k) + c\,{\mathscr T}({\lambda}_k) + d\,{\mathscr U}({\lambda}_k). \label{Yo10}
\end{equation}
As Bell notes \cite{Bell-1966}, it is not possible to work out the result of the measurement of a sum of non-commuting observables by trivially adding the results of separate measurements on each observable because each requires a distinct experiment.

The example Bell gives to illustrate this well known fact is that of the spin components of a fermion. A measurement of $\sigma_x$ can be made with a suitably oriented Stern-Gerlach magnet. But the measurement of $\sigma_y$ would require a different orientation of the magnet. And the measurement of the sum $(\sigma_x+\sigma_y)$ would again require a third and quite a different orientation of the magnet from the previous two. Consequently, the result of the last measurement --- {\it i.e.}, an eigenvalue of $(\sigma_x+\sigma_y)$ --- will not be the sum of an eigenvalue of $\sigma_x$ plus an eigenvalue of $\sigma_y$. The additivity of the expectation values, namely, $\langle\,\psi\,|\,\sigma_x\,|\,\psi\,\rangle+\langle\,\psi\,|\,\sigma_y\,|\,\psi\,\rangle=\langle\,\psi\,|\,\sigma_x+\,\sigma_y\,|\,\psi\,\rangle$, is a peculiar property of the quantum states $|\,\psi\,\rangle$. It would not hold for individual eigenvalues of non-commuting observables in a dispersion-free state of a hidden variable theory. In a dispersion-free state, every observable would have a unique value equal to one of its eigenvalues. And since there can be no linear relationship between the eigenvalues of non-commuting observables, the additivity relation (\ref{QM}) that holds for quantum mechanical states would not hold for dispersion-free states. Therefore, von Neumann's assumption (\ref{CM}) above --- as innocent and inevitable it may seem mathematically --- is wrong physically. And since at least one of the\break assumptions used in the proof of his theorem is wrong, his conclusion ruling out hidden variable theories is not valid.

It may appear that this objection by Bell holds only for individual eigenvalues of non-commuting observables and therefore Einstein's observation concerning the additivity of expectation values quoted above is not quite correct. In other words, it may appear that the oversight in von Neumann's theorem is that it not only assumes the additivity of expectation values but also additivity in every instance of eigenvalues. With the notation $v(A)$ for the eigenvalue of the observable $A$, this interpretation of the oversight in von Neumann's theorem is stated by Mermin as follows: ``But to require that $v(A+B)= v(A)+v(B)$ in each individual system of the ensemble is to ensure that a relation holds in the mean by imposing it case by case --- a sufficient, but hardly a necessary condition'' \cite{Mermin1993}. This interpretation, however, as popularized by Mermin, is not quite correct. Einstein's observation above that the additivity of expectation values need not hold for hidden variable theories even ``in the mean'' remains valid, as we have proved in Appendix \ref{A} below. As a result, because of the inequality (\ref{Yo10}), for non-commuting observables the quantum mechanical expectation value
\begin{equation}
\left\langle\,\psi\left|\left\{a\,{\mathcal R} + b\,{\mathcal S} + c\,{\mathcal T} + d\;{\mathcal U\,}\right\}\right|\psi\,\right\rangle \not=
\frac{1}{n}\sum_{k\,=\,1}^{n}
\big\{a\,{\mathscr R}({\lambda}_k) + b\,{\mathscr S}({\lambda}_k) + c\,{\mathscr T}({\lambda}_k) + d\,{\mathscr U}({\lambda}_k)\big\}. \label{non-exp}
\end{equation}
That is, the quantum mechanical expectation value of the operator ${a\,{\mathcal R} + b\,{\mathcal S} + c\,{\mathcal T} + d\;{\mathcal U}}$ is not equal to the average over the linear sum of the individual eigenvalues ${\mathscr R}({\lambda}_k)$, ${\mathscr S}({\lambda}_k)$, ${\mathscr T}({\lambda}_k)$, and ${\mathscr U}({\lambda}_k)$. In the Appendix \ref{A} below we have proved the inequality (\ref{non-exp}) and the fact that the eigenvalue of the sum ${a\,{\mathcal R} + b\,{\mathcal S} + c\,{\mathcal T} + d\;{\mathcal U}}$ of operators is not the sum $a\,{\mathscr R} + b\,{\mathscr S} + c\,{\mathscr T} + d\,{\mathscr U}$ of eigenvalues, unless ${\mathcal R}$, ${\mathcal S}$, ${\mathcal T}$, and ${\mathcal U}$ commute with each other. We have also proved that the inequality in (\ref{non-exp}) can reduce to equality {\it if and only if} $\,$the operators ${\mathcal R}$, ${\mathcal S}$, ${\mathcal T}$, and ${\mathcal U}$ commute with each other. But that means that, for non-commuting observables, in a dispersion-free state, assumption (\ref{CM}) is not correct:
\begin{equation}
\frac{a}{n}\sum_{k\,=\,1}^{n}
{\mathscr R}({\lambda}_k)+
\frac{b}{n}\sum_{k\,=\,1}^{n}
{\mathscr S}({\lambda}_k)+
\frac{c}{n}\sum_{k\,=\,1}^{n}
{\mathscr T}({\lambda}_k)+
\frac{d}{n}\sum_{k\,=\,1}^{n}
{\mathscr U}({\lambda}_k)\not=
\frac{1}{n}\sum_{k\,=\,1}^{n}
\big\{a\,{\mathscr R}({\lambda}_k) + b\,{\mathscr S}({\lambda}_k) + c\,{\mathscr T}({\lambda}_k) + d\,{\mathscr U}({\lambda}_k)\big\}.
\end{equation}

Incidentally, in the past decade Bub \cite{Bub} and Dieks \cite{Dieks} have attempted to revive von Neumann's theorem despite the fact that the existence since 1952 of Bohm's non-local hidden variable theory \cite{Bohm} provides a constructive refutation of the theorem, independently of its formal defect discussed above. As Mermin and Schack \cite{Mermin2018} in their refutation of the argument by Bub \cite{Bub} and Dieks \cite{Dieks} point out, the claim made by the latter authors amounts to insisting that von Neumann's assumption (\ref{CM}) is ``analytic'' and therefore perfectly valid. But that misses the point of the objection to the theorem raised by Einstein, Bell, and others, which has to do with physics, not mathematics. In particular, the identification (\ref{CM}) goes against the very definition of what is meant by a hidden variable theory. The notion of realism, as it was understood by Einstein, demands that every observable of a given physical system must possess a definite value as a preexisting property of the system, whether or not it is observed ({\it e.g.}, the moon exists, regardless of whether anyone is looking at it). Therefore, the first requirement of any hidden variable theory is simultaneous assignment of definite values to {\it all} observables of a given physical system. And as we saw above, these definite values must be the eigenvalues of the corresponding quantum mechanical operators. Now there are five observables involved in the quantum mechanical relation (\ref{QM}): $\mathcal{R}$, $\mathcal{S}$, $\mathcal{T}$, $\mathcal{U}$, and $\mathcal{X}=a\,\mathcal{R}+b\,\mathcal{S}+c\,\mathcal{T}+d\,\mathcal{U}$. Therefore, its hidden variable counterpart (\ref{CM}) must involve their respective eigenvalues ${\mathscr R}({\lambda}_k)$, ${\mathscr S}({\lambda}_k)$, ${\mathscr T}({\lambda}_k)$, ${\mathscr U}({\lambda}_k)$,
and ${\mathscr X}({\lambda}_k)$. But the linear sum of eigenvalues $a\,{\mathscr R}({\lambda}_k) + b\,{\mathscr S}({\lambda}_k) + c\,{\mathscr T}({\lambda}_k) + d\,{\mathscr U}({\lambda}_k)$ appearing on the RHS of (\ref{CM}) is not one of the eigenvalues ${\mathscr X}({\lambda}_k)$ of the operator $\mathcal{X}$,
\begin{equation}
a\,{\mathscr R}({\lambda}_k) + b\,{\mathscr S}({\lambda}_k) + c\,{\mathscr T}({\lambda}_k) + d\,{\mathscr U}({\lambda}_k)\not={\mathscr X}({\lambda}_k), \label{ls}
\end{equation}
when $\mathcal{R}$, $\mathcal{S}$, $\mathcal{T}$, and $\mathcal{U}$ are non-commuting, and this is the problem with the identification (\ref{CM}). Consequently, the claim by Bub \cite{Bub} and Dieks \cite{Dieks} that (\ref{CM}) is ``analytic'' and therefore it must be valid misses the point of the objection raised by Einstein, Bell, and others. In fact, given a definite eigenvalue ${\mathscr X}({\lambda}_k)$ of the operator $\mathcal{X}=a\,\mathcal{R}+b\,\mathcal{S}+c\,\mathcal{T}+d\,\mathcal{U}$, the hidden variable or dispersion-free counterpart of the quantum mechanical relation (\ref{QM}) is {\it not}, in general, the identification (\ref{CM}) that follows analytically, but necessarily the following physically correct and meaningful equality:
\begin{equation}
\boxed{\frac{a}{n}\sum_{k\,=\,1}^{n}
{\mathscr R}({\lambda}_k)+
\frac{b}{n}\sum_{k\,=\,1}^{n}
{\mathscr S}({\lambda}_k)+
\frac{c}{n}\sum_{k\,=\,1}^{n}
{\mathscr T}({\lambda}_k)+
\frac{d}{n}\sum_{k\,=\,1}^{n}
{\mathscr U}({\lambda}_k)=
\frac{1}{n}\sum_{k\,=\,1}^{n}
{\mathscr X}({\lambda}_k).} \label{14}
\end{equation}
It is this equality that must hold for hidden variable theories, regardless of their specific characteristics, {\it not} equality$\;$(\ref{CM}).

\section{The Oversight in Bell's Theorem}\label{III}

We now turn to Bell's theorem \cite{Bell-1964} and show that it harbours the same oversight as that in von Neumann's theorem. As we noted in the Introduction, Bell's theorem purports to prove that no locally causal and realistic theory in the sense envisaged by Einstein can reproduce all of the statistical predictions of quantum theory. We will follow the proof of the theorem in the manner of Clauser, Horne, Shimony, and Holt (CHSH) \cite{CHSH,Clauser}. In the derivation of the bounds on the CHSH correlator, such as those in Eq.~(\ref{chshbounds}) below, one usually employs factorized probabilities of observing the measurement results rather than the measurement results themselves as we will do in our derivation. But employing probabilities in that manner only manages to mask the implicit assumption in the proof we intend to bring out here.

Consider the standard EPR type spin-${\frac{1}{2}}$ experiment, as proposed by Bohm and later used by Bell in the proof of his theorem \cite{Bell-1964,Clauser}. Alice is free to choose a detector direction ${\bf a}$ or ${\bf a'}$ and Bob is free to choose a detector direction ${\bf b}$ or ${\bf b'}$ to detect spins of the fermions they receive from a common source, at a space-like distance from each other. The objects of interest then are the bounds on the sum of counterfactually possible averages of their observations put together in the manner of CHSH \cite{CHSH},
\begin{equation}
{\cal E}({\bf a},\,{\bf b})\,+\,{\cal E}({\bf a},\,{\bf b'})\,+\,{\cal E}({\bf a'},\,{\bf b})\,-\,{\cal E}({\bf a'},\,{\bf b'})\,, \label{CHSH}
\end{equation}
with each average, or expectation value, or correlation between the observations of Alice and Bob, defined as
\begin{equation}
{\cal E}({\bf a},\,{\bf b})\,=\lim_{\,n\,\gg\,1}\left[\frac{1}{n}\sum_{k\,=\,1}^{n}\,{\mathscr A}({\bf a},\,{\lambda}_k)\;{\mathscr B}({\bf b},\,{\lambda}_k)\right]\,\equiv\,\Bigl\langle\,{\mathscr A}_{k}({\bf a})\,{\mathscr B}_{k}({\bf b})\,\Bigr\rangle\,,\label{expect}
\end{equation}
where ${\mathscr A({\bf a},\,{\lambda}_k)\equiv {\mathscr A}_{k}({\bf a})=\pm1}$ and ${\mathscr B({\bf b},\,{\lambda}_k)\equiv {\mathscr B}_{k}({\bf b})=\pm1}$ are the respective measurement results of Alice and Bob. Note that ${\mathscr A({\bf a},\,{\lambda})}$ and ${\mathscr B({\bf b},\,{\lambda})}$ are manifestly local and realistic functions, where, following Bell\cite{Bell-1964}, locality is defined as follows:
\begin{quote}
\underbar{Local Causality}: Apart from the hidden variable ${\lambda}$, the result ${{\mathscr A}=\pm1}$ depends {\it only} on the measurement direction ${\bf a}$, chosen freely by Alice, regardless of Bob's actions. And, similarly, apart from the hidden variable ${\lambda}$, the result ${{\mathscr B}=\pm1}$ depends {\it only} on the measurement direction ${\bf b}$, chosen freely by Bob, regardless of Alice's actions. In particular, the function ${{\mathscr A}({\bf a},\,\lambda)}$ {\it does not} depend on ${\bf b}$ or ${\mathscr B}$ and the function ${{\mathscr B}({\bf b},\,\lambda)}$ {\it does not} depend on ${\bf a}$ or ${\mathscr A}$. Moreover, the hidden variable ${\lambda}$ does not depend on ${\bf a}$, ${\bf b}$, ${\mathscr A}$, or ${\mathscr B}$.
\end{quote}
Now, to derive the bound on the CHSH correlator (\ref{CHSH}), suppose that four independent experiments are carried out (each on a different day, for example), resulting in the following four separate averages, which can subsequently added together as
\begin{equation}
\Bigl\langle\,{\mathscr A}_{k_1}({\bf a})\,{\mathscr B}_{k_1}({\bf b})\,\Bigr\rangle\,+\, \Bigl\langle\,{\mathscr A}_{k_2}({\bf a})\,{\mathscr B}_{k_2}({\bf b'})\,\Bigr\rangle\,+\,\Bigl\langle\,{\mathscr A}_{k_3}({\bf a'})\,{\mathscr B}_{k_3}({\bf b})\,\Bigr\rangle\,-\, \Bigl\langle\,{\mathscr A}_{k_4}({\bf a'})\,{\mathscr B}_{k_4}({\bf b'})\,\Bigr\rangle, \label{8-1}
\end{equation}
where $k_i=k_1$, $k_2$, $k_3$, and $k_4$ are unrelated indices for each average. But since ${{\mathscr A}_{k_i}({\bf a})=\pm1}$ and ${{\mathscr B}_{k_i}({\bf b})=\pm1}$, the average of the product of the measurement results in each case would be ${-1\leqslant\Bigl\langle\,{\mathscr A}_{k_i}({\bf a})\,{\mathscr B}_{k_i}({\bf b})\,\Bigr\rangle\leqslant +1}$. Consequently, we can immediately read off the upper and lower bounds on the sequence of four averages as
\begin{equation}
-4\,\leqslant\,\Bigl\langle\,{\mathscr A}_{k_1}({\bf a})\,{\mathscr B}_{k_1}({\bf b})\,\Bigr\rangle\,+\, \Bigl\langle\,{\mathscr A}_{k_2}({\bf a})\,{\mathscr B}_{k_2}({\bf b'})\,\Bigr\rangle\,+\,\Bigl\langle\,{\mathscr A}_{k_3}({\bf a'})\,{\mathscr B}_{k_3}({\bf b})\,\Bigr\rangle\,-\, \Bigl\langle\,{\mathscr A}_{k_4}({\bf a'})\,{\mathscr B}_{k_4}({\bf b'})\,\Bigr\rangle\,\leqslant\,+4\,. \label{8-2}
\end{equation}
Next, the assumption of local realism allows Bell to consider counterfactual possibilities such as the following: Alice could have chosen the detector direction ${\bf a}'$ instead of ${\bf a}$ and thus observed the result ${\mathscr A}_{k_1}({\bf a}')$ instead of ${\mathscr A}_{k_1}({\bf a})$ while Bob observed the result ${\mathscr B}_{k_1}({\bf b})$ choosing the detector direction ${\bf b}$, for the same run of the experiment with the same pair of particles represented by the index $k_1$. Such counterfactual possibilities, together with the assumption of a large number of trials for each experiment, allow him to equate all four indices $k_1$, $k_2$, $k_3$, and $k_4$ to the same index $k=k_1=k_2=k_3=k_4$, reducing the sequence in (\ref{8-2}) to
\begin{equation}
-4\,\leqslant\,\Bigl\langle\,{\mathscr A}_{k}({\bf a})\,{\mathscr B}_{k}({\bf b})\,\Bigr\rangle\,+\, \Bigl\langle\,{\mathscr A}_{k}({\bf a})\,{\mathscr B}_{k}({\bf b'})\,\Bigr\rangle\,+\,\Bigl\langle\,{\mathscr A}_{k}({\bf a'})\,{\mathscr B}_{k}({\bf b})\,\Bigr\rangle\,-\, \Bigl\langle\,{\mathscr A}_{k}({\bf a'})\,{\mathscr B}_{k}({\bf b'})\,\Bigr\rangle\,\leqslant\,+4\,, \label{9-1}
\end{equation}
giving the bounds of $\pm4$ on the CHSH correlator (\ref{CHSH}), thanks to the definition of the correlation (\ref{expect}):
\begin{equation}
-4\,\leqslant\,{\cal E}({\bf a},\,{\bf b})\,+\,{\cal E}({\bf a},\,{\bf b'})\,+\,{\cal E}({\bf a'},\,{\bf b})\,-\,{\cal E}({\bf a'},\,{\bf b'})\,\leqslant\,+4\,. \label{711}
\end{equation}
At this stage we can question, on physical grounds, the significance of the identification of the four independent set of experiments, characterized by the indices $k_1$, $k_2$, $k_3$, and $k_4$, with a single set of experiments involving counterfactually possible detector directions, characterized by the same index $k=k_1=k_2=k_3=k_4$. Because, physically, nothing has changed between the experiments considered in (\ref{8-2}) and (\ref{9-1}), apart from the increased number $n$ of trials for each set \cite{CHSH,Clauser}. The experiments considered in (\ref{9-1}) are still four entirely independent sets of experiments that may have been carried out on four different days, for example. But we need not question the identification $k=k_1=k_2=k_3=k_4$ to recognize that there is no reason so far to believe that, after (\ref{9-1}), the bounds on the CHSH correlator (\ref{CHSH}) are anyway tighter than $\pm4$. So far, the assumptions of (a) local realism and (b) a large number of trials for each counterfactually possible detector direction have led us to the bounds in (\ref{711}) on the CHSH correlator that are no tighter$\;$than$\;\pm4$.

This should have been Bell's final conclusion. However, by continuing, Bell overlooked something that is physically unjustifiable. By assuming the {\it additivity of expectation values}, without which his theorem {\it cannot} be formally proven, Bell identified the above sum of four separate averages of real numbers (to be recorded, for example, in experimental runs performed on four different days) with the following single average:
\begin{equation}
{\cal E}({\bf a},\,{\bf b})\,+\,{\cal E}({\bf a},\,{\bf b'})\,+\,{\cal E}({\bf a'},\,{\bf b})\,-\,{\cal E}({\bf a'},\,{\bf b'})\,=\,\Bigl\langle\,{\mathscr A}_{k}({\bf a})\,{\mathscr B}_{k}({\bf b})\,+\,{\mathscr A}_{k}({\bf a})\,{\mathscr B}_{k}({\bf b'})\,+\,{\mathscr A}_{k}({\bf a'})\,{\mathscr B}_{k}({\bf b})\,-\,{\mathscr A}_{k}({\bf a'})\,{\mathscr B}_{k}({\bf b'})\,\Bigr\rangle\,. \label{rep}
\end{equation}
As innocuous as this step may seem mathematically, it is in fact an illegitimate step physically, because what is being averaged on its RHS are unobservable and unphysical quantities. Moreover, it is not dictated by the assumptions of local realism and a large number of trials for each counterfactually possible experiment, both of which have been exhausted by the stage (\ref{711}) of the derivation. It is a mathematical step, analogous to von Neumann's assumption (\ref{CM}), quite independent of the assumptions of local realism and a large number of trials. But while it has no counterpart in physics, it allows Bell to reduce the sum in (\ref{9-1}) to
\begin{equation}
\Bigl\langle\,{\mathscr A}_{k}({\bf a})\,\big\{\,{\mathscr B}_{k}({\bf b})+{\mathscr B}_{k}({\bf b'})\,\big\}\,+\,{\mathscr A}_{k}({\bf a'})\,\big\{\,{\mathscr B}_{k}({\bf b})-{\mathscr B}_{k}({\bf b'})\,\big\}\,\Bigr\rangle\,.\label{absurd}
\end{equation}
And since ${{\mathscr B}_{k}({\bf b})=\pm1}$, if ${|{\mathscr B}_{k}({\bf b})+{\mathscr B}_{k}({\bf b'})|=2}$, then ${|{\mathscr B}_{k}({\bf b})-{\mathscr B}_{k}({\bf b'})|=0}$, and vice versa. Consequently, using ${{\mathscr A}_{k}({\bf a})=\pm1}$, it is easy to conclude that the absolute value of the above average cannot exceed 2, just as Bell concluded: 
\begin{equation}
-2\,\leqslant\,\Bigl\langle\,{\mathscr A}_{k}({\bf a})\,{\mathscr B}_{k}({\bf b})\,+\,
{\mathscr A}_{k}({\bf a})\,{\mathscr B}_{k}({\bf b'})\,+\,{\mathscr A}_{k}({\bf a'})\,{\mathscr B}_{k}({\bf b})\,-\,{\mathscr A}_{k}({\bf a'})\,{\mathscr B}_{k}({\bf b'})\,\Bigr\rangle\,\leqslant\,+2\,.\label{5}
\end{equation}
Given the identification (\ref{rep}) (the LHS of which is a perfectly meaningful physical quantity), the last calculation then allows Bell to conclude that the CHSH correlator (\ref{CHSH}) is bounded by $\pm2$, thereby facilitating the experimental tests of local realism:
\begin{equation}
-2\,\leqslant\,{\cal E}({\bf a},\,{\bf b})\,+\,{\cal E}({\bf a},\,{\bf b'})\,+\,{\cal E}({\bf a'},\,{\bf b})\,-\,{\cal E}({\bf a'},\,{\bf b'})\,\leqslant\,+2\,. \label{chshbounds}
\end{equation}
In summary, the stringent bounds of $\pm2$ on the CHSH correlator derived by Bell follow from three assumptions: (a) local realism, (b) a large number of trials for each counterfactually possible experiment, and (c) the identification represented by Eq.~(\ref{rep}); namely, the additivity of expectation values: {\it a sum of expectation values is equal to the expectation value of the sum}.

Sometimes it is argued that the last assumption --- assumption (c) --- is not an additional assumption. It is argued that the assumption of local realism and the appeal to statistics makes the addition of expectation values in Eq.~(\ref{rep}) possible. It is argued that local realism makes the individual expectation values take definite values on the real line of numbers, which can then be added to give the inequalities (\ref{chshbounds}). But this argument belies the fact that an additional step, namely (\ref{rep}), over and above the assumption of local realism, is inevitable in any formal proof of Bell's theorem. It is not automatically implied by local realism and the appeal to statistics. It must be acknowledged as an additional, if implicit assumption. An appeal to the large-$n$ limit to make the LHS of Eq.~(\ref{rep}) tend to its RHS is nothing but assumption (c) in disguise. It is true that for large $n$ the LHS of Eq.~(\ref{rep}) tends to its RHS making that identification possible. But that is only the first step. It brings the horse to the waterhole, but does not make it drink the water. 

Let us now try to understand why the identification in Eq.~(\ref{rep}) is illegitimate. To begin with, Einstein's, or even Bell's own notion of local-realism, does not, by itself, demand this identification. Since this notion is captured already in the very definition of the functions ${{\mathscr A}({\bf a},\,{\lambda}_k)}$ \cite{Bell-1964}, together with the procedure by which we have arrived at the averages in (\ref{9-1}), the LHS of the identification (\ref{rep}) satisfies the demand of local realism perfectly well. Nor can the statistical equivalence between (\ref{8-2}) and (\ref{9-1}) we have granted to arrive at the LHS of (\ref{rep}) justify its replacement with the single average on its RHS, {\it at the expense of what is physically possible in the actual experiments}. To be sure, just as mathematically there is nothing wrong with von Neumann's physically unjustified identification (\ref{CM}), mathematically there is nothing wrong with the identification of four separate averages in (\ref{9-1}) with a single average in (\ref{absurd}). Indeed, every schoolchild knows that a sum of averages is equal to the average of the sum. But this rule of thumb is not valid in the above case, because ${({\bf a},\,{\bf b})}$, ${({\bf a},\,{\bf b'})}$, ${({\bf a'},\,{\bf b})}$, and ${({\bf a'},\,{\bf b'})}$ are {\it mutually exclusive pairs of measurement directions}, corresponding to four {\it incompatible} experiments. Each pair can be used by Alice and Bob for a given experiment, for all runs ${1}$ to ${n}$, but no two of the four pairs can be used by them simultaneously. This is because Alice and Bob do not have the ability to make measurements along counterfactually possible pairs of directions such as ${({\bf a},\,{\bf b})}$ and ${({\bf a},\,{\bf b'})}$ simultaneously. Alice, for example, can make measurements along ${\bf a}$ or ${\bf a'}$, but not along ${\bf a}$ {\it and} ${\bf a'}$ at the same time.

But this inconvenient fact is rather devastating for Bell's argument, because it means that his identification (\ref{rep}) is illegitimate. Consider a specific run of the EPR-Bohm experiment we started out with and the quantity being averaged in (\ref{rep}):
\begin{equation}
{\mathscr A}_{k}({\bf a})\,{\mathscr B}_{k}({\bf b})\,+\,
{\mathscr A}_{k}({\bf a})\,{\mathscr B}_{k}({\bf b'})\,+\,{\mathscr A}_{k}({\bf a'})\,{\mathscr B}_{k}({\bf b})\,-\,
{\mathscr A}_{k}({\bf a'})\,{\mathscr B}_{k}({\bf b'})\,. \label{riy}
\end{equation}
Here the index ${k=1}$ now represents a specific run of the experiment. But since Alice and Bob have only two particles at their disposal for each run, only one of the four terms of the above sum is physically meaningful. In other words, the above quantity in (\ref{riy}) is physically meaningless, because Alice, for example, cannot align her detector along ${\bf a}$ and ${\bf a'}$ at the same time. And likewise, Bob cannot align his detector along ${\bf b}$ and ${\bf b'}$ at the same time. What is more, this will be true for all possible runs of the experiment, or equivalently for all possible pairs of particles. But that, in turn, implies that all of the quantities listed below, as they appear in the average (\ref{5}), are unobservable, and hence physically meaningless:
\begin{align}
&{\mathscr A}_{1}({\bf a})\,{\mathscr B}_{1}({\bf b})\,+\,
{\mathscr A}_{1}({\bf a})\,{\mathscr B}_{1}({\bf b'})\,+\,{\mathscr A}_{1}({\bf a'})\,{\mathscr B}_{1}({\bf b})\,-\,
{\mathscr A}_{1}({\bf a'})\,{\mathscr B}_{1}({\bf b'})\,, \nonumber \\
&{\mathscr A}_{2}({\bf a})\,{\mathscr B}_{2}({\bf b})\,+\,
{\mathscr A}_{2}({\bf a})\,{\mathscr B}_{2}({\bf b'})\,+\,{\mathscr A}_{2}({\bf a'})\,{\mathscr B}_{2}({\bf b})\,-\,
{\mathscr A}_{2}({\bf a'})\,{\mathscr B}_{2}({\bf b'})\,,  \nonumber \\
&{\mathscr A}_{3}({\bf a})\,{\mathscr B}_{3}({\bf b})\,+\,
{\mathscr A}_{3}({\bf a})\,{\mathscr B}_{3}({\bf b'})\,+\,{\mathscr A}_{3}({\bf a'})\,{\mathscr B}_{3}({\bf b})\,-\,
{\mathscr A}_{3}({\bf a'})\,{\mathscr B}_{3}({\bf b'})\,,  \nonumber \\
&{\mathscr A}_{4}({\bf a})\,{\mathscr B}_{4}({\bf b})\,+\,
{\mathscr A}_{4}({\bf a})\,{\mathscr B}_{4}({\bf b'})\,+\,{\mathscr A}_{4}({\bf a'})\,{\mathscr B}_{4}({\bf b})\,-\,
{\mathscr A}_{4}({\bf a'})\,{\mathscr B}_{4}({\bf b'})\,,  \nonumber \\
&\;\;\;\;\boldsymbol{\cdot}\nonumber \\
&\;\;\;\;\boldsymbol{\cdot}\nonumber \\
&\;\;\;\;\boldsymbol{\cdot}\nonumber \\
&\!{\mathscr A}_{n}({\bf a})\,{\mathscr B}_{n}({\bf b})\,+\,
{\mathscr A}_{n}({\bf a})\,{\mathscr B}_{n}({\bf b'})\,+\,{\mathscr A}_{n}({\bf a'})\,{\mathscr B}_{n}({\bf b})\,-\,
{\mathscr A}_{n}({\bf a'})\,{\mathscr B}_{n}({\bf b'})\,.\nonumber
\end{align}
But since each of the quantities above is physically meaningless, their average appearing on the RHS of (\ref{rep}), namely
\begin{equation}
\Bigl\langle\,{\mathscr A}_{k}({\bf a})\,{\mathscr B}_{k}({\bf b})\,+\,{\mathscr A}_{k}({\bf a})\,{\mathscr B}_{k}({\bf b'})\,+\,{\mathscr A}_{k}({\bf a'})\,{\mathscr B}_{k}({\bf b})\,-\,{\mathscr A}_{k}({\bf a'})\,{\mathscr B}_{k}({\bf b'})\,\Bigr\rangle\,, \label{abave}
\end{equation}
is also physically meaningless. That is to say, no physical experiment can be performed --- {\it even in principle} --- that can meaningfully allow to measure or evaluate the above average, since none of the above list of quantities could have experimentally observable values. Therefore, the innocuous looking identification (\ref{rep}) of Bell is, in fact, illegitimate.

In the extensive literature on Bell's theorem the identification (\ref{rep}) is often justified on the grounds of ``counterfactual definiteness'', a variant of realism. But, as we will see in more detail at the end of the next section, such a justification also does not hold water. It is usually argued in this regard that, because each of the four results appearing in the summation being averaged in (\ref{abave}) is at least counterfactually realizable, the quantity (\ref{riy}) representing their sum is also counterfactually meaningful. However, this argument is mistaken. While each of the four results appearing in the summation being averaged in (\ref{abave}) is indeed realizable  counterfactually, the quantity (\ref{riy}) representing their sum cannot be realized {\it even} counterfactually, precisely because only one of the four results can be realized actually in a given run of the experiment. But this implies that, because the LHS of the identification (\ref{rep}) is a sum of actually realizable averages, the identification (\ref{rep}) amounts to asserting
\begin{equation}
\text{physically meaningful quantity} = \text{physically meaningless quantity}. \label{absurd-2}
\end{equation}
Moreover, as noted in the paragraph after Eq.~(\ref{chshbounds}), it is not possible to prove Bell's theorem formally without taking the step (\ref{rep}). Without this step, the horse is brought to the waterhole but it fails to drink the water --- {\it i.e.}, without this step, the bounds on the CHSH correlator cannot be tighter than $\pm4$. But the absurdity of the equality (\ref{absurd-2}) suggests\break that local realism is not implemented correctly in the proof of Bell's theorem. It compels us to search for a physical quantity that is at least counterfactually realizable so that the RHS of Eq.~(\ref{rep}) can be corrected, as we shall do next. 

\section{Comparing the oversights in the theorems of von Neumann and Bell} \label{IV}

Now, to compare Bell's oversight we just discussed with von Neumann's oversight discussed in Section \ref{II}, let us view the measurement result ${\mathscr A}({\bf a},\,{\lambda}_k)\,{\mathscr B}({\bf b},\,{\lambda}_k)=+1$ or $-1$, simultaneously observed by Alice and Bob at remote stations, as a specific eigenvalue ${\mathscr R}({\bf a},\,{\bf b},\,{\lambda}_k)$ of the observable ${\mathcal R}({\bf a},\,{\bf b})\equiv{\boldsymbol\sigma}_1\cdot{\bf a}\,\otimes\,{\boldsymbol\sigma}_2\cdot{\bf b}$ in the quantum mechanical singlet state
\begin{equation}
|\psi_{\bf n}\rangle=\frac{1}{\sqrt{2}}\Bigl\{|{\bf n},\,+\rangle_1\otimes
|{\bf n},\,-\rangle_2\,-\,|{\bf n},\,-\rangle_1\otimes|{\bf n},\,+\rangle_2\Bigr\}\,.
\label{single}
\end{equation}
Here ${\bf n}$ is an arbitrary unit direction in space, subscripts 1 and 2 refer to particles 1 and 2, and the eigenvalue equation
\begin{equation}
{\boldsymbol\sigma}\cdot{\bf n}\,|{\bf n},\,\pm\rangle\,=\,
\pm\,|{\bf n},\,\pm\rangle \label{spin}
\end{equation}
determines the quantum mechanical eigenstates in which the particles have spin ``up'' or ``down'' in the units of ${\hslash=2}$, with ${\boldsymbol\sigma}$ being the Pauli spin ``vector'' ${({\sigma_x},\,{\sigma_y},\,{\sigma_z})}$. As before, our interest lies in comparing the quantum predictions of spin correlations between the constituent fermions with those derived within some locally causal dispersion-free theory \cite{Bell-1964}. For that purpose, we make the following identifications between the notations used in Sections \ref{II} and \ref{III}:
\begin{align}
{\mathscr A}({\bf a},\,{\lambda}_k)\,{\mathscr B}({\bf b},\,{\lambda}_k)&\equiv {\mathscr R}({\bf a},\,{\bf b},\,{\lambda}_k)=\pm1\;\;\text{is an eigenvalue of the observable}\;\;{\mathcal R}({\bf a},\,{\bf b})\equiv{\boldsymbol\sigma}_1\cdot{\bf a}\,\otimes\,{\boldsymbol\sigma}_2\cdot{\bf b} \label{ab} \\
{\mathscr A}({\bf a},\,{\lambda}_k)\,{\mathscr B}({\bf b'},\,{\lambda}_k)&\equiv {\mathscr S}({\bf a},\,{\bf b'},\,{\lambda}_k)=\pm1\;\;\text{is an eigenvalue of the observable}\;\;{\mathcal S}({\bf a},\,{\bf b'})\equiv{\boldsymbol\sigma}_1\cdot{\bf a}\,\otimes\,{\boldsymbol\sigma}_2\cdot{\bf b'} \\
{\mathscr A}({\bf a'},\,{\lambda}_k)\,{\mathscr B}({\bf b},\,{\lambda}_k)&\equiv {\mathscr T}({\bf a'},\,{\bf b},\,{\lambda}_k)=\pm1\;\;\text{is an eigenvalue of the observable}\;\;{\mathcal T}({\bf a'},\,{\bf b})\equiv{\boldsymbol\sigma}_1\cdot{\bf a'}\,\otimes\,{\boldsymbol\sigma}_2\cdot{\bf b} \\
{\mathscr A}({\bf a'},\,{\lambda}_k)\,{\mathscr B}({\bf b'},\,{\lambda}_k)&\equiv {\mathscr U}({\bf a'},\,{\bf b'},\,{\lambda}_k)=\pm1\;\;\text{is an eigenvalue of the observable}\;\;{\mathcal U}({\bf a'},\,{\bf b'})\equiv{\boldsymbol\sigma}_1\cdot{\bf a'}\,\otimes\,{\boldsymbol\sigma}_2\cdot{\bf b'}. \label{a'b'}
\end{align}
With these identifications we can now rewrite the CHSH correlator (\ref{CHSH}) for $n{\gg}1$ as a sum of the average values as 
\begin{equation}
{\cal E}({\bf a},\,{\bf b})\,+\,{\cal E}({\bf a},\,{\bf b'})\,+\,{\cal E}({\bf a'},\,{\bf b})\,-\,{\cal E}({\bf a'},\,{\bf b'})\equiv\frac{1}{n}\sum_{k\,=\,1}^{n}
{\mathscr R}({\lambda}_k)+
\frac{1}{n}\sum_{k\,=\,1}^{n}
{\mathscr S}({\lambda}_k)+
\frac{1}{n}\sum_{k\,=\,1}^{n}
{\mathscr T}({\lambda}_k)-
\frac{1}{n}\sum_{k\,=\,1}^{n}
{\mathscr U}({\lambda}_k) \label{equal}
\end{equation}
by setting the scalar coefficients appearing in (\ref{QM}) and (\ref{CM}) to $a=b=c=+1$ and $d=-1$, and, similarly, the RHS of Eq.~(\ref{rep}) as
\begin{equation}
\Bigl\langle\,{\mathscr A}_{k}({\bf a})\,{\mathscr B}_{k}({\bf b})\,+\,{\mathscr A}_{k}({\bf a})\,{\mathscr B}_{k}({\bf b'})\,+\,{\mathscr A}_{k}({\bf a'})\,{\mathscr B}_{k}({\bf b})\,-\,{\mathscr A}_{k}({\bf a'})\,{\mathscr B}_{k}({\bf b'})\,\Bigr\rangle\equiv\frac{1}{n}\sum_{k\,=\,1}^{n}\big\{{\mathscr R}({\lambda}_k) + {\mathscr S}({\lambda}_k) + {\mathscr T}({\lambda}_k) - {\mathscr U}({\lambda}_k)\big\},
\end{equation}
so that Bell's assumption (\ref{rep}) can be rewritten as
\begin{equation}
\frac{1}{n}\sum_{k\,=\,1}^{n}
{\mathscr R}({\lambda}_k)+
\frac{1}{n}\sum_{k\,=\,1}^{n}
{\mathscr S}({\lambda}_k)+
\frac{1}{n}\sum_{k\,=\,1}^{n}
{\mathscr T}({\lambda}_k)-
\frac{1}{n}\sum_{k\,=\,1}^{n}
{\mathscr U}({\lambda}_k)=
\frac{1}{n}\sum_{k\,=\,1}^{n}
\big\{{\mathscr R}({\lambda}_k) + {\mathscr S}({\lambda}_k) + {\mathscr T}({\lambda}_k) - {\mathscr U}({\lambda}_k)\big\}. \label{EPRB-2}
\end{equation}

Now, as we discussed in the previous two sections, mathematically this assumption --- without which the stringent bounds of $-2$ and $+2$ on the CHSH correlator cannot be derived --- is a trivial identity. But is it meaningful physically? To answer this question, note that the spin observables ${\mathcal R}({\bf a},\,{\bf b})$, ${\mathcal S}({\bf a},\,{\bf b'})$, ${\mathcal T}({\bf a'},\,{\bf b})$, and ${\mathcal U}({\bf a'},\,{\bf b'})$ defined in (\ref{ab}) to (\ref{a'b'}) do not commute with each other \cite{Bell-1966}. Consequently, in parallel with the discussion in Section \ref{II}, in the dispersion-free counterpart $|\psi_{\bf n},\,\lambda\,\rangle$ of the state (\ref{single}) the result of the measurement of their sum ${\mathcal R} + {\mathcal S} + {\mathcal T} - {\mathcal U}$, or the eigenvalue of the observable ${\mathcal R} + {\mathcal S} + {\mathcal T} - {\mathcal U\,}$, will not be the sum ${\mathscr R}({\lambda}_k) + {\mathscr S}({\lambda}_k) + {\mathscr T}({\lambda}_k) - {\mathscr U}({\lambda}_k)$ of the individual eigenvalues:
\begin{equation}
\left\langle\,\psi_{\bf n},\,\lambda_k\left|\left\{{\mathcal R} + {\mathcal S} + {\mathcal T} - {\mathcal U\,}\right\}\right|\psi_{\bf n},\,\lambda_k\,\right\rangle \not=
{\mathscr R}({\lambda}_k) + {\mathscr S}({\lambda}_k) + {\mathscr T}({\lambda}_k) - {\mathscr U}({\lambda}_k),
\end{equation}
and therefore --- as proved in Appendix \ref{A} for arbitrary $a$, $b$, $c$, and $d$ --- the quantum mechanical expectation value
\begin{equation}
\left\langle\,\psi_{\bf n}\left|\left\{{\mathcal R} + {\mathcal S} + {\mathcal T} - {\mathcal U\,}\right\}\right|\psi_{\bf n}\,\right\rangle \not=
\frac{1}{n}\sum_{k\,=\,1}^{n}
\big\{{\mathscr R}({\lambda}_k) + {\mathscr S}({\lambda}_k) + {\mathscr T}({\lambda}_k) - {\mathscr U}({\lambda}_k)\big\}. \label{inequal}
\end{equation}
But in the light of (\ref{QM}), (\ref{equal}), and the identifications of the observables in (\ref{ab}) to (\ref{a'b'}), this lack of equivalence implies
\begin{equation}
\frac{1}{n}\sum_{k\,=\,1}^{n}
{\mathscr R}({\lambda}_k)+
\frac{1}{n}\sum_{k\,=\,1}^{n}
{\mathscr S}({\lambda}_k)+
\frac{1}{n}\sum_{k\,=\,1}^{n}
{\mathscr T}({\lambda}_k)-
\frac{1}{n}\sum_{k\,=\,1}^{n}
{\mathscr U}({\lambda}_k)\not=
\frac{1}{n}\sum_{k\,=\,1}^{n}
\big\{{\mathscr R}({\lambda}_k) + {\mathscr S}({\lambda}_k) + {\mathscr T}({\lambda}_k) - {\mathscr U}({\lambda}_k)\big\}. \label{EPRB}
\end{equation}
Consequently, the assumption (\ref{rep}), or equivalently the assumption (\ref{EPRB-2}), is physically wrong. As a result, the stringent bounds of $\pm2$ on the CHSH correlator derived on the basis of this assumption are also wrong. It is not at all surprising that they are violated in the Bell-test experiments \cite{Aspect}.  It is also important to note that in arriving at this conclusion we have not compromised either locality or realism. We have merely brought out a flaw in the logic of Bell's argument.

In summary, since Bell's theorem, above all, is a theorem against a class of hidden variable theories, it is no different, in this regard, from von Neumann's theorem \cite{Shimony}. Consequently, it too pertains to simultaneous (albeit contextual \cite{Bell-1966}) assignment of definite values to {\it all} of the observables of a relevant quantum system, thereby making the corresponding hidden variable theory compatible with realism \cite{EPR}, or counterfactual definiteness \cite{Clauser}. The notion of realism demands that every observable of a given physical system must possess a definite value as a preexisting property, whether or not it is actually observed (``the moon exists whether or not it is observed''). Therefore, the first requirement of any hidden variable theory, whether it is local or nonlocal, is simultaneous assignment of definite values to all observables of a given physical system. But as we saw in the previous section, realism is not implemented correctly in the proof of Bell’s theorem. For it to be implemented correctly, the quantity being averaged on the right-hand side of Eq.~(\ref{rep}) would have to be realizable at least counterfactually. Now there are five observables associated with the Hilbert space of the singlet systems investigated in the Bell-test experiments; namely, $\mathcal{R}$, $\mathcal{S}$, $\mathcal{T}$, $\mathcal{U}$, and 
\begin{equation}
{\mathcal X}:={\mathcal R}\,+\,{\mathcal S}\,+\,{\mathcal T}-\,{\mathcal U}\equiv{\boldsymbol\sigma}_1\cdot{\bf a}\,\otimes\,\{{\boldsymbol\sigma}_2\cdot{\bf b}\,+\,{\boldsymbol\sigma}_2\cdot{\bf b'}\}\,+\,{\boldsymbol\sigma}_1\cdot{\bf a'}\,\otimes\,\{{\boldsymbol\sigma}_2\cdot{\bf b}\,-\,{\boldsymbol\sigma}_2\cdot{\bf b'}\}. \label{F}
\end{equation}
Among these, $\mathcal{X}$ is never observed in the experiments we considered in Section \ref{II}. But it does exist as a self-adjoint operator in the corresponding quantum theory. According to von Neumann's Hilbert space formulation of quantum mechanics \cite{vonNeumann,Albertson,Ballentine}, a sum of self-adjoint operators such as ${\mathcal X}={\mathcal R}\,+\,{\mathcal S}\,+\,{\mathcal T}-\,{\mathcal U}$ is also a self-adjoint operator in the corresponding Hilbert space. Moreover, the correspondence between self-adjoint operators and observables is one-to-one \cite{vonNeumann,Ballentine,Mermin2018}. Therefore, realism demands that all five of the above observables must possess preexisting values, whether of not they are observed. And within any hidden variable theory these values must be one of the eigenvalues of the corresponding quantum mechanical operators, as explained in Section~\ref{II}. Now Bell's theorem, perhaps unintentionally, does assign a definite value to ${\mathcal X}$, but the value it ends up assigning is
\begin{equation}
{\mathscr R}({\lambda}_k) + {\mathscr S}({\lambda}_k) + {\mathscr T}({\lambda}_k) - {\mathscr U}({\lambda}_k)\equiv{\mathscr A}_{k}({\bf a})\,{\mathscr B}_{k}({\bf b})\,+\,{\mathscr A}_{k}({\bf a})\,{\mathscr B}_{k}({\bf b'})\,+\,{\mathscr A}_{k}({\bf a'})\,{\mathscr B}_{k}({\bf b})\,-\,{\mathscr A}_{k}({\bf a'})\,{\mathscr B}_{k}({\bf b'}), \label{notcoreig}
\end{equation}
which is not the correct eigenvalue of ${\mathcal X}$. As we have derived in the Appendix \ref{A} below, the correct eigenvalue of ${\mathcal X}$ is
\begin{equation}
{\mathscr X}({\lambda}_k) 
= \sqrt{\big\{\,{\mathscr R}({\lambda}_k) + {\mathscr S}({\lambda}_k) + {\mathscr T}({\lambda}_k) - {\mathscr U}({\lambda}_k)\big\}^2 +\, \langle\,\psi\,|\,{\mathcal Y}\,|\,\psi\,\rangle\,}, \label{coreig}
\end{equation}
which is a highly nonlinear function of $\langle\,\psi\,|\,{\mathcal Y}\,|\,\psi\,\rangle$ and the linear sum $\big\{\,{\mathscr R}({\lambda}_k) + {\mathscr S}({\lambda}_k) + {\mathscr T}({\lambda}_k) - {\mathscr U}({\lambda}_k)\big\}$ of eigenvalues ${\mathscr R}({\lambda}_k)$, ${\mathscr S}({\lambda}_k)$, ${\mathscr T}({\lambda}_k)$, and $-{\mathscr U}({\lambda}_k)$, necessitating the inequality (\ref{inequal}), as we have proven in Appendix \ref{A}. More importantly, ${\mathscr X}({\lambda}_k)$ is at least a counterfactually realizable result, because it is a {\it bona fide} eigenvalue of ${\mathcal X}$. Therefore, the corrected version of the identification (\ref{EPRB-2}), with the correct implementation of local realism, would be via the following replacement on its RHS: 
\begin{equation}
{\mathscr R}({\lambda}_k) + {\mathscr S}({\lambda}_k) + {\mathscr T}({\lambda}_k) - {\mathscr U}({\lambda}_k)\longrightarrow\sqrt{\big\{\,{\mathscr R}({\lambda}_k) + {\mathscr S}({\lambda}_k) + {\mathscr T}({\lambda}_k) - {\mathscr U}({\lambda}_k)\big\}^2 +\, \langle\,\psi\,|\,{\mathcal Y}\,|\,\psi\,\rangle\,}. \label{ls-3}
\end{equation}
But since $\langle\,\psi\,|\,{\mathcal Y}\,|\,\psi\,\rangle\not=0$ in general, the absolute bounds on the CHSH correlator (\ref{CHSH}) would then exceed $2$ in general.

It is evident from the above discussion that the physically unjustified assumption (\ref{EPRB-2}) Bell's theorem depends on is homologous to the additivity assumption (\ref{CM}) von Neumann's theorem depends on. This conclusion, however, holds only for non-commuting observables. In exact analogy with von Neumann's assumption (\ref{CM}), the RHS of the identification (\ref{EPRB-2}) is physically meaningless only for non-commuting observables. Consequently, it may be tempting to conclude that Bell's theorem is valid but only excludes hidden variable theories corresponding to commuting observables. However, the relevant quantum mechanical observables do not commute even for the singlet state. Therefore, the trivial case of commuting observables is physically uninteresting. On the other hand, when the relevant observables do commute, the eigenvalue (\ref{coreig}) reduces to (\ref{notcoreig}), thereby recovering the incorrect implementation of local realism posited by Bell. 

\section{Correct Local-realistic Bounds on the CHSH Correlator}

In the previous section we saw that
${\mathscr R}({\lambda}_k) + {\mathscr S}({\lambda}_k) + {\mathscr T}({\lambda}_k) - {\mathscr U}({\lambda}_k)$ is {\it not} one of the eigenvalues of the observable ${\mathcal X}$, and therefore it cannot be used to work out the correct local-realistic bounds on the CHSH correlator. Indeed, as we saw in Section~\ref{III}, the extreme values of ${\mathscr R}({\lambda}_k) + {\mathscr S}({\lambda}_k) + {\mathscr T}({\lambda}_k) - {\mathscr U}({\lambda}_k)$ are $-2$ and $+2$, leading to the incorrect bounds (\ref{chshbounds}) on the CHSH correlator (\ref{CHSH}) because of the physical incompatibility between (\ref{QM}) and (\ref{CM}). Fortunately, the correct local-realistic bounds on the CHSH correlator can be easily worked out by working out the extrema of the expectation (or average) value of the operator ${\mathcal X}$ defined in (\ref{F}), and they work out to be ${-2\sqrt{2}}$ and ${+2\sqrt{2}}$ as follows:

Let ${\mathscr X}({\lambda}_k)\not={\mathscr R}({\lambda}_k) + {\mathscr S}({\lambda}_k) + {\mathscr T}({\lambda}_k) - {\mathscr U}({\lambda}_k)$ be an eigenvalue of the observable  ${\mathcal X}$ specified in (\ref{F}). In Section~\ref{III} we saw that ${\mathcal X}$ is not observable in a typical EPR-Bohm experiment involving only two particles per run. However, as a self-adjoint operator on the Hilbert space of the composite system, it is an observable in theory, according to von Neumann's Hilbert space formulation of quantum mechanics. Therefore, regardless of whether ${\mathcal X}$ is actually measured in the experiment, in the dispersion-free counterpart $|\,\psi_{\bf n},\,\lambda\,\rangle$ of $|\,\psi_{\bf n}\,\rangle$ we would have
\begin{equation}
\left\langle\,\psi_{\bf n},\,\lambda_k\,|\,{\mathcal X}\,|\,\psi_{\bf n},\,\lambda_k\,\right\rangle =
{\mathscr X}({\lambda}_k). \label{34}
\end{equation}
Consequently, the quantum mechanical expectation value of ${\mathcal X}$ can be recovered by uniform averaging over $\lambda$ as before:
\begin{equation}
\left\langle\,\psi_{\bf n}\left|\,{\mathcal X}\,\right|\psi_{\bf n}\,\right\rangle =
\frac{1}{n}\sum_{k\,=\,1}^{n}
\left\langle\,\psi_{\bf n},\,\lambda_k\,|\,{\mathcal X}\,|\,\psi_{\bf n},\,\lambda_k\,\right\rangle
= \frac{1}{n}\sum_{k\,=\,1}^{n}
{\mathscr X}({\lambda}_k). \label{35}
\end{equation}
Using (\ref{35}), the relation (\ref{EPRB-2}) can then be corrected to give the following relation that is immune to a Bell-type criticism:
\begin{equation}
\frac{1}{n}\sum_{k\,=\,1}^{n}
{\mathscr R}({\lambda}_k)+
\frac{1}{n}\sum_{k\,=\,1}^{n}
{\mathscr S}({\lambda}_k)+
\frac{1}{n}\sum_{k\,=\,1}^{n}
{\mathscr T}({\lambda}_k)-
\frac{1}{n}\sum_{k\,=\,1}^{n}
{\mathscr U}({\lambda}_k)=
\frac{1}{n}\sum_{k\,=\,1}^{n}
{\mathscr X}({\lambda}_k), \label{EPRB-3}
\end{equation}
which is a special case of the general relation (\ref{14}) for hidden variable theories with the scalar coefficients appearing in (\ref{QM}) and (\ref{CM}) set to $a=b=c=+1$ and $d=-1$. Unlike (\ref{EPRB-2}), this is the correct dispersion-free counterpart of the quantum mechanical relation (\ref{QM}) among the operators ${\mathcal R}$, ${\mathcal S}$, ${\mathcal T}$, ${\mathcal U}$, and ${\mathcal X}$. Unlike that between (\ref{QM}) and (\ref{CM}) brought out by Bell \cite{Bell-1966}, there is no physical incompatibility between the linear relationship (\ref{QM}) among the quantum mechanical expectation values and its dispersion-free counterpart (\ref{EPRB-3}).

Now, one way to derive the correct bounds on the CHSH correlator (\ref{CHSH}) is by recalling from our discussion leading to Eq.~(\ref{chshbounds}) that $-2\leqslant {\mathscr R}({\lambda}_k) + {\mathscr S}({\lambda}_k) + {\mathscr T}({\lambda}_k) - {\mathscr U}({\lambda}_k)\leqslant+2\,$ and by establishing that $-4\leqslant\langle\,\psi\,|\,{\mathcal Y}\,|\,\psi\,\rangle\leqslant+4$, so that the eigenvalue ${\mathscr X}(\lambda_k)$ expressed in (\ref{coreig}) can be deduced to be bounded by $\pm2\sqrt{2}$. Since these bounds would hold for each $\lambda_k$, the CHSH correlator (\ref{CHSH}), or equivalently the LHS of Eq.~(\ref{EPRB-3}), would also be bounded by $-2\sqrt{2}$ and $+2\sqrt{2}$. However, it is cumbersome to establish that $-4\leqslant\langle\,\psi\,|\,{\mathcal Y}\,|\,\psi\,\rangle\leqslant+4$. Therefore, we will employ a simpler method to derive these bounds:

Note that local commutativity of the operators acting on the spacelike separated subsystems $1$ and $2$ dictates that
\begin{equation}
\left[\,{\boldsymbol\sigma}_1\cdot{\bf a},\, {\boldsymbol\sigma}_2\cdot{\bf b}\,\right] = 0\;\;\;\forall\;\;{\bf a}\;\;\text{and}\;\;{\bf b}. \label{37}
\end{equation}
Moreover, we have the identities 
\begin{equation}
({\boldsymbol\sigma}_1\cdot{\bf a})^2=({\boldsymbol\sigma}_1\cdot{\bf a'})^2=({\boldsymbol\sigma}_2\cdot{\bf b})^2=({\boldsymbol\sigma}_2\cdot{\bf b'})^2 = \dbl, \label{38}
\end{equation}
where $\dbl$ is a $2\times2$ identity matrix. For the next steps, it is convenient to introduce the following Tsirel'son operator, 
\begin{equation}
{\mathcal Z}:=\frac{1}{\sqrt{2}}\left( {\boldsymbol\sigma}_1\cdot{\bf a}-\frac{\left({\boldsymbol\sigma}_2\cdot{\bf b}+{\boldsymbol\sigma}_2\cdot{\bf b'}\right)}{\sqrt{2}}\right)^{\!2}
+\,\frac{1}{\sqrt{2}}\left( {\boldsymbol\sigma}_1\cdot{\bf a'}-\frac{\left({\boldsymbol\sigma}_2\cdot{\bf b}-{\boldsymbol\sigma}_2\cdot{\bf b'}\right)}{\sqrt{2}}\right)^{\!2}\!, \label{Tsirel}
\end{equation}
which is a sum of squared Hermitian operators and therefore its expectation value in any state would be non-negative:
\begin{equation}
\langle\,\psi_{\bf n}\left|\,{\mathcal Z}\,\right|\psi_{\bf n}\,\rangle \geqslant 0. \label{40}
\end{equation}
By expanding  the RHS of Eq.~(\ref{Tsirel}) and using the relations (\ref{37}) and (\ref{38}), the Tsirel'son operator can be simplified to  
\begin{equation}
{\mathcal Z}=2\sqrt{2}\,\dbl - {\mathcal X}.
\end{equation}
Then, using the non-negativity (\ref{40}) of $\langle\,{\mathcal Z}\,\rangle$, it is easy to see that 
\begin{equation}
\langle\,\psi_{\bf n}\left|\,{\mathcal Z}\,\right|\psi_{\bf n}\,\rangle=2\sqrt{2} -
\langle\,\psi_{\bf n}\left|\,{\mathcal X}\,\right|\psi_{\bf n}\,\rangle \geqslant 0.
\end{equation}
Consequently, we have the desired bounds:
\begin{equation}
-2\sqrt{2}\,\leqslant
\langle\,\psi_{\bf n}\left|\,{\mathcal X}\,\right|\psi_{\bf n}\,\rangle \leqslant +2\sqrt{2}.
\end{equation}

These are the extrema of the quantum mechanical expectation value $\langle\,\psi_{\bf n}\left|\,{\mathcal X}\,\right|\psi_{\bf n}\,\rangle$ of the operator defined in (\ref{F}). However, according to the relations (\ref{35}) and (\ref{EPRB-3}), the above inequalities are equivalent to the inequalities
\begin{equation}
-2\sqrt{2} \,\leqslant \left\{\frac{1}{n}\sum_{k\,=\,1}^{n}
{\mathscr R}({\lambda}_k)+
\frac{1}{n}\sum_{k\,=\,1}^{n}
{\mathscr S}({\lambda}_k)+
\frac{1}{n}\sum_{k\,=\,1}^{n}
{\mathscr T}({\lambda}_k)-
\frac{1}{n}\sum_{k\,=\,1}^{n}
{\mathscr U}({\lambda}_k)\right\} \leqslant +2\sqrt{2}, \label{Q-bounds}
\end{equation}
giving the correct bounds on the CHSH correlator (\ref{CHSH}). In the notation of (\ref{CHSH}), they can be expressed as
\begin{equation}
-2\sqrt{2} \,\leqslant \,{\cal E}({\bf a},\,{\bf b})\,+\,{\cal E}({\bf a},\,{\bf b'})\,+\,{\cal E}({\bf a'},\,{\bf b})\,-\,{\cal E}({\bf a'},\,{\bf b'}) \leqslant +2\sqrt{2}. \label{chshqm}
\end{equation}
We have thus derived the quantum mechanical Tsirel'son bounds \cite{Tsirelson} in a purely local-realistic, dispersion-free setting. Needless to add, these corrected bounds on the CHSH correlator have never been violated in any Bell-test experiments. 
Therefore, far from ruling out local realism, the experimental violations of the stringent bounds of $\pm2$ on the CHSH correlator (\ref{CHSH}) derived in Section~\ref{II} rule out Bell's assumption of the additivity of expectation values represented by Eq.~(\ref{rep}), or equivalently by Eq.~(\ref{EPRB-2}). 
\begin{figure}
\centering
\includegraphics[scale=1]{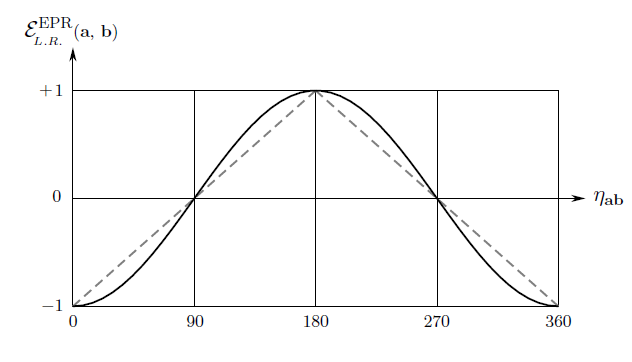}
\caption{A graph of the singlet correlations predicted by the local-realistic 3-sphere model presented in \cite{IJTP,RSOS,IEEE-Access,Socks}. The x-axis depicts the angle in degrees between the vectors ${\bf a}$ and ${\bf b}$ freely chosen by Alice and Bob and the y-axis depicts the value of the correlations. The dotted straight lines represent classical correlations predicted by Bell's theorem for any local-realistic model.}
\label{fig}
\end{figure}

Not surprisingly, the above conclusion is further supported by the independent derivations in references \cite{RSOS,IEEE-Access,Socks} of the bounds derived in (\ref{chshqm}) within a manifestly local-realistic 3-sphere model \cite{Disproof,IJTP} (cf. Fig.~\ref{fig}):
\begin{equation}
-2\sqrt{2}\,\leqslant\,{\cal E}({\bf a},\,{\bf b})\,+\,{\cal E}({\bf a},\,{\bf b'})\,+\,{\cal E}({\bf a'},\,{\bf b})\,-\,{\cal E}({\bf a'},\,{\bf b'})\,=\,-\,\cos\eta_{{\bf a}{\bf b}}\,-\,\cos\eta_{{\bf a}{\bf b'}}\,-\,\cos\eta_{{\bf a'}{\bf b}}\,+\,\cos\eta_{{\bf a'}{\bf b'}}\,\leqslant\,2\sqrt{2}\,,
\end{equation}
where $\eta_{{\bf a}{\bf b}}$, $\eta_{{\bf a}{\bf b'}}$, $\eta_{{\bf a'}{\bf b}}$, and $\eta_{{\bf a'}{\bf b'}}$ are the angles between respective detector directions Alice and Bob can choose counterfactually. The above inequalities can be easily verified by considering, for example, the following extreme values of the angles: $\eta_{{\bf a}{\bf b}}=45^\circ$, $\eta_{{\bf a}{\bf b'}}=45^\circ$, $\eta_{{\bf a'}{\bf b}}=45^\circ$, and $\eta_{{\bf a'}{\bf b'}}=135^\circ$. In this model the results ${{\mathscr A}({\bf a},\,\lambda)=\pm1}$ and ${{\mathscr B}({\bf b},\,\lambda)=\pm1}$ observed remotely by Alice and Bob are taken to be limiting scalar points of a quaternionic 3-sphere, $S^3$, taken as a model for the three-dimensional physical space, instead of its usual model as ${\mathrm{I\!R}^3}$, and the correlations ${\cal E}({\bf a},\,{\bf b})$ between their results are calculated using the prescription
\begin{equation}
{\cal E}^{\rm EPR}_{_{\!{L.R.}}}({\bf a},\,{\bf b})\,=\lim_{\,n\,\gg\,1}\left[\frac{1}{n}\sum_{k\,=\,1}^{n}\,{\mathscr A}({\bf a},\,{\lambda}^k)\;{\mathscr B}({\bf b},\,{\lambda}^k)\right]=-\,\cos\eta_{{\bf a}{\bf b}}\,. \label{prepcorr}
\end{equation}
Needless to say, this local-realistic model provides a constructive counterexample to Bell's theorem, complementing the critique of its formal aspects presented in the current paper. Its upshot is that the strong correlations we observe in Nature have little to do with quantum entanglement {\it per se}. Rather, they are consequences of the compact geometry and topology of the physical space we live in; namely, of the quaternionic 3-sphere \cite{IEEE-Access,Socks}. In fact, any hidden variable\break model based on the functions ${{\mathscr A}({\bf a},\,\lambda)=\pm1}$ and ${{\mathscr B}({\bf b},\,\lambda)=\pm1}$ respecting the conditions of local causality specified in Section \ref{III} that reproduces the singlet correlations calculated using the above prescription would have the desired local-realistic properties. There would be no need to consider or investigate any Bell-type inequalities for this purpose.

\section{Concluding remarks}

We have compared the mistaken assumptions in the respective theorems of von Neumann and Bell against hidden variable theories and found them to be homologous. While Bell's theorem concerns a rather restricted class of locally causal theories, both theorems claim to rule out theories based on dispersion-free states by assuming the relationship
\begin{equation}
\frac{a}{n}\sum_{k\,=\,1}^{n}
{\mathscr R}({\lambda}_k)+
\frac{b}{n}\sum_{k\,=\,1}^{n}
{\mathscr S}({\lambda}_k)+
\frac{c}{n}\sum_{k\,=\,1}^{n}
{\mathscr T}({\lambda}_k)+
\frac{d}{n}\sum_{k\,=\,1}^{n}
{\mathscr U}({\lambda}_k)=
\frac{1}{n}\sum_{k\,=\,1}^{n}
\big\{a\,{\mathscr R}({\lambda}_k) + b\,{\mathscr S}({\lambda}_k) + c\,{\mathscr T}({\lambda}_k) + d\,{\mathscr U}({\lambda}_k)\big\} \tag{\ref{CM}}
\end{equation}
between expected values of observables to arrive at their respective conclusions. But, rather ironically, it is precisely the requirement of hypothetical dispersion-free states that forces the above assumption to fail for hidden variable theories. It is the very notion of realism --- the idea that the outcomes of individual measurement events are predetermined, with simultaneous assignment of definite values to all observables of a physical system --- that forces the assumption (\ref{CM}) to\break fail, as Bell himself has argued in his criticism of von Neumann's theorem \cite{Bell-1966}. In Section 3 of his paper Bell writes:
\begin{quote}
Thus the formal proof of von Neumann does not justify his informal conclusion. ... It was not the objective measurable predictions of quantum mechanics which ruled out hidden variables. It was the arbitrary assumption of a particular (and impossible) relation between the results of incompatible measurements either of which might be made on a given occasion but only one of which can in fact be made \cite{Bell-1966}.
\end{quote}
But that is precisely the mistake Bell himself has made in his own famous theorem \cite{Bell-1964,CHSH}. Bell's theorem can be proven only by considering three or four incompatible physical experiments involving mutually exclusive detector directions. Since experiments along mutually exclusive detector directions cannot be performed simultaneously, they amount to observing non-commuting observables. And for non-commuting observables of a system in a dispersion-free state, the result of the measurement of a sum of the observables is not the same as the sum of the results of the measurements of the individual observables in that sum \cite{Bell-1966}. Therefore, Bell's assumption (\ref{rep}), or equivalently (\ref{EPRB-2}), does not hold for dispersion-free states, rendering the informal conclusion of the formal proof of his theorem invalid. Paraphrasing Bell from the above quote, we can therefore conclude: It is not the objectively measurable predictions of quantum mechanics that rule out the possibility of a local-realistic theory. It is the {\it ad hoc} assumption of three or four physically incompatible experiments, any one of which might be performed on a given occasion, but only one of which can, in fact, be performed in practice. 

Fortunately, Bell's oversight can be rectified by replacing the equality (\ref{EPRB-2}) with the physically meaningful equality
\begin{equation}
\frac{1}{n}\sum_{k\,=\,1}^{n}
{\mathscr R}({\lambda}_k)+
\frac{1}{n}\sum_{k\,=\,1}^{n}
{\mathscr S}({\lambda}_k)+
\frac{1}{n}\sum_{k\,=\,1}^{n}
{\mathscr T}({\lambda}_k)-
\frac{1}{n}\sum_{k\,=\,1}^{n}
{\mathscr U}({\lambda}_k)=
\frac{1}{n}\sum_{k\,=\,1}^{n}
{\mathscr X}({\lambda}_k), \tag{46}
\end{equation}
where ${\mathscr X}({\lambda}_k)$, unlike ${\mathscr R}({\lambda}_k) + {\mathscr S}({\lambda}_k) + {\mathscr T}({\lambda}_k) - {\mathscr U}({\lambda}_k)$, is one of the {\it bona fide} eigenvalues of the self-adjoint operator $\mathcal{X}=\mathcal{R}+\mathcal{S}+\mathcal{T}-\mathcal{U}$. Consequently, unlike (\ref{rep}) or (\ref{EPRB-2}), this equality correctly implements local realism, which requires simultaneous (albeit contextual) assignment of definite values to all observables of a physical system. Once this\break correction is made to (\ref{EPRB-2}), the bounds on the CHSH correlator work out to be ${\pm2\sqrt{2}}$ instead of ${\pm2}$, thereby mitigating the conclusion of Bell's theorem. Consequently, what is ruled out by Bell-test experiments \cite{Clauser,Aspect} is not local realism, but the additivity of expectation values, (\ref{rep}) or (\ref{EPRB-2}), which does not hold for hidden variable theories to begin with.

\appendix 

\section{Eigenvalue of a Sum of Non-Commuting Operators and the Proof of Eq.~(\ref{non-exp})} \label{A}

In this appendix we prove that the eigenvalue of the sum ${a\,{\mathcal R} + b\,{\mathcal S} + c\,{\mathcal T} + d\;{\mathcal U}}$ of operators is not equal to the sum $a\,{\mathscr R} + b\,{\mathscr S} + c\,{\mathscr T} + d\,{\mathscr U}$ of the individual eigenvalues of the operators ${\mathcal R}$, ${\mathcal S}$, ${\mathcal T}$, and ${\mathcal U}$, unless they commute with each other. It is not difficult to prove this by first evaluating the square of the operator ${\{a\,{\mathcal R} + b\,{\mathcal S} + c\,{\mathcal T} + d\;{\mathcal U}\}}$ as follows:
\begin{align}
\{a\,{\mathcal R} + b\,{\mathcal S} + c\,{\mathcal T} + d\;{\mathcal U}\}\{a\,{\mathcal R} + b\,{\mathcal S} + c\,{\mathcal T} + d\;{\mathcal U}\} &= a^2{\mathcal R}^2 + ab\,{\mathcal R}{\mathcal S}+ ac\,{\mathcal R}{\mathcal T}+ ad\,{\mathcal R}{\mathcal U} \notag \\
&\;\;\;\;\;\;\;\;\;\;+ ba\,{\mathcal S}{\mathcal R}+ b^2 {\mathcal S}^2+ bc\,{\mathcal S}{\mathcal T}+ bd\,{\mathcal S}{\mathcal U} \notag \\
&\;\;\;\;\;\;\;\;\;\;+ ca\,{\mathcal T}{\mathcal R}+ cb\,{\mathcal T}{\mathcal S}+ c^2 {\mathcal T}^2+ cd\,{\mathcal T}{\mathcal U} \notag \\
&\;\;\;\;\;\;\;\;\;\;+ da\,{\mathcal U}{\mathcal R}+ db\,{\mathcal U}{\mathcal S}+ dc\,{\mathcal U}{\mathcal T}+ d^2 {\mathcal U}^2. \label{square}
\end{align}
Now, assuming that the operators ${\mathcal R}$, ${\mathcal S}$, ${\mathcal T}$, and ${\mathcal U}$ do not commute in general, let us define the following operators:
\begin{align}
{\mathcal L}&:={\mathcal S}{\mathcal R}-{\mathcal R}{\mathcal S} \;\Longleftrightarrow\; {\mathcal S}{\mathcal R} =  {\mathcal R}{\mathcal S}+{\mathcal L}, \label {def-i}\\
{\mathcal M}&:={\mathcal T}{\mathcal R}-{\mathcal R}{\mathcal T} \;\Longleftrightarrow\; {\mathcal T}{\mathcal R} =  {\mathcal R}{\mathcal T}+{\mathcal M}, \\
{\mathcal N}&:={\mathcal T}{\mathcal S}-{\mathcal S}{\mathcal T} \;\Longleftrightarrow\; {\mathcal T}{\mathcal S} =  {\mathcal S}{\mathcal T}+{\mathcal N}, \\
{\mathcal O}&:={\mathcal U}{\mathcal R}-{\mathcal R}{\mathcal U} \;\Longleftrightarrow\; {\mathcal U}{\mathcal R} =  {\mathcal R}{\mathcal U}+{\mathcal O}, \\
{\mathcal P}&:={\mathcal U}{\mathcal T}-{\mathcal T}{\mathcal U} \;\Longleftrightarrow\; {\mathcal U}{\mathcal T} =  {\mathcal T}{\mathcal U}+{\mathcal P}, \\
\text{and}\;\;\;{\mathcal Q}&:={\mathcal U}{\mathcal S}-{\mathcal S}{\mathcal U} \;\Longleftrightarrow\; {\mathcal U}{\mathcal S} =  {\mathcal S}{\mathcal U}+{\mathcal Q}. \label{def-f}
\end{align}
These operators would be null operators with vanishing eigenvalues if the operators ${\mathcal R}$, ${\mathcal S}$, ${\mathcal T}$, and ${\mathcal U}$ did commute with each other. Using the above relations for the operators ${\mathcal S}{\mathcal R}$, ${\mathcal T}{\mathcal R}$, ${\mathcal T}{\mathcal S}$, ${\mathcal U}{\mathcal R}$, ${\mathcal U}{\mathcal T}$ and ${\mathcal U}{\mathcal S}$, Eq.~(\ref{square}) can be simplified to
\begin{align}
\{a\,{\mathcal R} + b\,{\mathcal S} + c\,{\mathcal T} + d\;{\mathcal U}\}\{a\,{\mathcal R} + b\,{\mathcal S} + c\,{\mathcal T} + d\;{\mathcal U}\} &= a^2{\mathcal R}^2 + 2ab\,{\mathcal R}{\mathcal S}+ 2ac\,{\mathcal R}{\mathcal T}+ 2ad\,{\mathcal R}{\mathcal U} \notag \\
&\;\;\;\;\;\;\;\;\;\;+ ab\,{\mathcal L}+ b^2 {\mathcal S}^2+ 2bc\,{\mathcal S}{\mathcal T}+ 2bd\,{\mathcal S}{\mathcal U} \notag \\
&\;\;\;\;\;\;\;\;\;\;+ ac\,{\mathcal M}+ bc\,{\mathcal N}+ c^2 {\mathcal T}^2+ 2cd\,{\mathcal T}{\mathcal U} \notag \\
&\;\;\;\;\;\;\;\;\;\;+ ad\,{\mathcal O}+ bd\,{\mathcal Q}+ cd\,{\mathcal P}+ d^2 {\mathcal U}^2 \\
&=\{a\,{\mathcal R} + b\,{\mathcal S} + c\,{\mathcal T} + d\;{\mathcal U}\}_{\bf c}^2\,+\,{\mathcal Y}, \label{A9}
\end{align}
\vspace{-0.4cm}
${\!\!}$where
\begin{equation}
{\mathcal Y}:=ab\,{\mathcal L} + ac\,{\mathcal M} + bc\,{\mathcal N}+ ad\,{\mathcal O}+ cd\,{\mathcal P} + bd\,{\mathcal Q}\,. \label{line}
\end{equation}
We have thus separated out the commuting part ${\{a\,{\mathcal R} + b\,{\mathcal S} + c\,{\mathcal T} + d\;{\mathcal U}\}_{\bf c}}$ and the non-commuting part ${\mathcal Y}$ of the operator ${\{a\,{\mathcal R} + b\,{\mathcal S} + c\,{\mathcal T} + d\;{\mathcal U}\}}$. Note that the operators ${\mathcal L}$, ${\mathcal M}$, ${\mathcal N}$, ${\mathcal O}$, ${\mathcal P}$, and ${\mathcal Q}$ defined in (\ref{def-i}) to (\ref{def-f}) will not commute with each other in general unless their constituents ${\mathcal R}$, ${\mathcal S}$, ${\mathcal T}$, and ${\mathcal U}$ themselves are commuting. Next, we work out the eigenvalue ${\mathscr X}$ of the operator ${{\mathcal X}:=\{a\,{\mathcal R} + b\,{\mathcal S} + c\,{\mathcal T} + d\;{\mathcal U}\,\}}$ in a state $|\,\psi\,\rangle$ using the eigenvalue equations
\begin{equation}
{\mathcal X}\,|\,\psi\,\rangle = {\mathscr X}\,|\,\psi\,\rangle 
\end{equation}
and
\begin{equation}
{\mathcal X}\,{\mathcal X}\,|\,\psi\,\rangle =
{\mathcal X}\big\{{\mathcal X}\,|\,\psi\,\rangle\big\} = {\mathcal X}\,\big\{{\mathscr X}\,|\,\psi\,\rangle\big\} =
{\mathscr X}\,\big\{{\mathcal X}\,|\,\psi\,\rangle\big\} =
{\mathscr X}^2\,|\,\psi\,\rangle, \label{A11}
\end{equation}
in terms of the eigenvalues ${\mathscr R}$, ${\mathscr S}$, ${\mathscr T}$, and ${\mathscr U}$ of the operators ${\mathcal R}$, ${\mathcal S}$, ${\mathcal T}$, and ${\mathcal U}$ and the expectation value $\langle\,\psi\,|\,{\mathcal Y}\,|\,\psi\,\rangle$:
\begin{equation}
{\mathscr X} = \sqrt{\langle\,\psi\,|\,{\mathcal X}\,{\mathcal X}\,|\,\psi\,\rangle}=\sqrt{\langle\,\psi\,|\big\{a\,{\mathcal R} + b\,{\mathcal S} + c\,{\mathcal T} + d\;{\mathcal U}\big\}_{\bf c}^2\,|\,\psi\,\rangle+\langle\,\psi\,|\,{\mathcal Y}\,|\,\psi\,\rangle\,},
\end{equation}
where we have used Eq.~(\ref{A9}). But the eigenvalue of the commuting part ${\{a\,{\mathcal R} + b\,{\mathcal S} + c\,{\mathcal T} + d\;{\mathcal U}\}_{\bf c}}$ is simply the linear sum ${a\,{\mathscr R} + b\,{\mathscr S} + c\,{\mathscr T} + d\,{\mathscr U}}$ of the eigenvalues of the operators ${\mathcal R}$, ${\mathcal S}$, ${\mathcal T}$, and ${\mathcal U}$. Consequently, using the equation analogous to (\ref{A11}) for the square of the operator ${\big\{a\,{\mathcal R} + b\,{\mathcal S} + c\,{\mathcal T} + d\;{\mathcal U}\big\}_{\bf c}}$ we can express the eigenvalue ${\mathscr X}$ of ${\mathcal X}$ as 
\begin{equation}
{\mathscr X}=\sqrt{\big\{a\,{\mathscr R} + b\,{\mathscr S} + c\,{\mathscr T} + d\,{\mathscr U}\big\}^2 + \,\langle\,\psi\,|\,{\mathcal Y}\,|\,\psi\,\rangle\,}.
\label{A12}
\end{equation}
Now, because the operators ${\mathcal L}$, ${\mathcal M}$, ${\mathcal N}$, ${\mathcal O}$, ${\mathcal P}$, and ${\mathcal Q}$ defined in Eqs.~(\ref{def-i}) to (\ref{def-f}) will not commute with each other in general if their constituent operators ${\mathcal R}$, ${\mathcal S}$, ${\mathcal T}$, and ${\mathcal U}$ are non-commuting, the eigenvalue ${\mathscr Y}$ of the operator ${\mathcal Y}$ defined in (\ref{line}) will not be equal to the linear sum of the corresponding eigenvalues ${\mathscr L}$, ${\mathscr M}$, ${\mathscr N}$, ${\mathscr O}$, ${\mathscr P}$, and ${\mathscr Q}$ in general,
\begin{equation}
{\mathscr Y}\not=ab\,{\mathscr L} + ac\,{\mathscr M} + bc\,{\mathscr N} + ad\,{\mathscr O} + cd\,{\mathscr P} + bd\,{\mathscr Q}\,,
\end{equation}
even if we assume that the operators ${\mathcal X}$ and ${\mathcal Y}$ commute with each other so that $\langle\,\psi\,|\,{\mathcal Y}\,|\,\psi\,\rangle={\mathscr Y}$ is an eigenvalue of ${\mathcal Y}$.\break That is to say, just like the eigenvalue ${\mathscr X}$ of ${\mathcal X}$, the eigenvalue ${\mathscr Y}$ of ${\mathcal Y}$ is also a nonlinear function in general. On the other hand, because we wish to prove that the eigenvalue of the sum ${a\,{\mathcal R} + b\,{\mathcal S} + c\,{\mathcal T} + d\;{\mathcal U}}$ of the operators ${\mathcal R}$, ${\mathcal S}$, ${\mathcal T}$, and ${\mathcal U}$ is not equal to the sum $a\,{\mathscr R} + b\,{\mathscr S} + c\,{\mathscr T} + d\,{\mathscr U}$ of the individual eigenvalues of the operators ${\mathcal R}$, ${\mathcal S}$, ${\mathcal T}$, and ${\mathcal U}$ unless they commute with each other, we must make sure that the eigenvalue ${\mathscr Y}$ does not vanish for the unlikely case in\break which the operators ${\mathcal L}$, ${\mathcal M}$, ${\mathcal N}$, ${\mathcal O}$, ${\mathcal P}$, and ${\mathcal Q}$ commute with each other. But even in that unlikely case we would have
\begin{equation}
{\mathscr Y}=ab\,{\mathscr L} + ac\,{\mathscr M} + bc\,{\mathscr N} + ad\,{\mathscr O} + cd\,{\mathscr P} + bd\,{\mathscr Q}\,
\end{equation}
as eigenvalue of the operator ${\mathcal Y}$ defined in (\ref{line}), and consequently the eigenvalue ${\mathscr X}$ in (\ref{A12}) will at best reduce to
\begin{equation}
{\mathscr X}=\sqrt{\big\{a\,{\mathscr R} + b\,{\mathscr S} + c\,{\mathscr T} + d\,{\mathscr U}\big\}^2 + ab\,{\mathscr L} + ac\,{\mathscr M} + bc\,{\mathscr N} + ad\,{\mathscr O} + cd\,{\mathscr P} + bd\,{\mathscr Q}\,}.
\end{equation}
In other words, even in such an unlikely case ${\mathscr Y}$ will not vanish, and consequently the eigenvalue ${\mathscr X}$ will not reduce$\;$to
\begin{equation}
{\mathscr X}=a\,{\mathscr R} + b\,{\mathscr S} + c\,{\mathscr T} + d\,{\mathscr U}. \label{A18}
\end{equation}
As a result, we can now prove Eq.~(\ref{non-exp}) of Section \ref{II}: Unless $\langle\,\psi\,|\,{\mathcal Y}\,|\,\psi\,\rangle\equiv0$, the average of the eigenvalue ${\mathscr X}$ will be
\begin{align}
\langle\,\psi\,|\,{\mathcal X}\,|\,\psi\,\rangle =
\frac{1}{n}\!\sum_{k\,=\,1}^{n}{\mathscr X}({\lambda}_k) 
&=\frac{1}{n}\!\sum_{k\,=\,1}^{n} \sqrt{\big\{a\,{\mathscr R}({\lambda}_k) + b\,{\mathscr S}({\lambda}_k) + c\,{\mathscr T}({\lambda}_k) + d\,{\mathscr U}({\lambda}_k)\big\}^2 + \,\langle\,\psi,\,\lambda_k\,|\,{\mathcal Y}\,|\,\psi,\,\lambda_k\,\rangle\,} \label{A19} \\
&\not=\frac{1}{n}\!\sum_{k\,=\,1}^{n} \sqrt{\big\{a\,{\mathscr R}({\lambda}_k) + b\,{\mathscr S}({\lambda}_k) + c\,{\mathscr T}({\lambda}_k) + d\,{\mathscr U}({\lambda}_k)\big\}^2 + \,{\mathscr Y}(\lambda_k)} \;\;\text{if}\;[\,{\mathcal X},\,{\mathcal Y}\,]\not=0\label{A20} \\
&\not= \frac{1}{n}\!\sum_{k\,=\,1}^{n}
\big\{a\,{\mathscr R}({\lambda}_k) + b\,{\mathscr S}({\lambda}_k) + c\,{\mathscr T}({\lambda}_k) + d\,{\mathscr U}({\lambda}_k)\big\}\;\text{if}\;{\mathscr L},\,{\mathscr M},\,{\mathscr N},\,{\mathscr O},\,{\mathscr P},\,{\mathscr Q}\not=0. \label{A21}
\end{align}
This proves Eq.~(11) discussed in Section \ref{II}. Note that, because ${\mathscr X}(\lambda_k)$ and $\langle\,\psi\,|\,{\mathcal Y}\,|\,\psi\,\rangle$ are highly {\it nonlinear} functions in general (recall, {\it e.g.}, that $\sqrt{x^2\pm y^2\,}\not=\sqrt{x^2}\,\pm\sqrt{y^2}\,$), the inequality in (\ref{A21}) can reduce to equality {\it if and only if} the operators ${\mathcal R}$, ${\mathcal S}$, ${\mathcal T}$, and ${\mathcal U}$ commute with each other. In that case, the operators ${\mathcal L}$, ${\mathcal M}$, ${\mathcal N}$, ${\mathcal O}$, ${\mathcal P}$, and ${\mathcal Q}$ defined in (\ref{def-i}) to (\ref{def-f}) will also commute with each other, as well as being null operators, with each of the eigenvalues ${\mathscr L}$, ${\mathscr M}$, ${\mathscr N}$, ${\mathscr O}$, ${\mathscr P}$, and ${\mathscr Q}$ reducing to zero. Consequently, in that case $\langle\,\psi\,|\,{\mathcal Y}\,|\,\psi\,\rangle$ will vanish identically and (\ref{A12}) will reduce to (\ref{A18}). In particular, for the CHSH correlator (\ref{equal}) for which ${a=b=c=+1}$ and ${d=-1}$, (\ref{A18}) will simplify to
\begin{equation}
{\mathscr X}(\lambda_k)={\mathscr R}(\lambda_k) + {\mathscr S}(\lambda_k) + {\mathscr T}(\lambda_k) - {\mathscr U}(\lambda_k). \label{Bell-e}
\end{equation}
It is {\it this} eigenvalue ${\mathscr X}(\lambda_k)$ in (\ref{Bell-e}) that has been implicitly and unjustifiably assumed in the proof of Bell's theorem.

\section*{Acknowledgements}

The similarity between von Neumann's oversight and Bell's oversight in their respective theorems brought out in this paper was first brought to my attention in passing by Michel Fodje, in November 2013, in a private correspondence.

\end{document}